\documentclass[twocolumn]{aastex63}

\graphicspath{{./}{figures/}}

\submitjournal{ApJ}

\shortauthors{Sharma et al.}

\begin{document}

\title{Constraining the Chemical Signatures and the Outburst Mechanism of the Class 0 Protostar HOPS 383}

\correspondingauthor{Rajeeb Sharma}
\email{rajeeb.sharma-1@ou.edu}

\author{Rajeeb Sharma}
\affil{Homer L. Dodge Department of Physics and Astronomy, University of Oklahoma, 440 W. Brooks Street, Norman, OK 73019, USA}

\author{John J. Tobin}
\affiliation{National Radio Astronomy Observatory, Charlottesville, VA 22901, USA}

\author{Patrick D. Sheehan}
\affiliation{National Radio Astronomy Observatory, Charlottesville, VA 22901, USA}

\author{S. Thomas Megeath}
\affiliation{Ritter Astrophysical Observatory, Department of Physics and Astronomy, University of Toledo, Toledo, OH, USA}

\author{William J. Fischer}
\affiliation{Space Telescope Science Institute, Baltimore, MD, USA}

\author{Jes K. J\o rgensen}
\affiliation{Niels Bohr Institute, University of Copenhagen, \O ster Voldgade 5-7, DK-1350, Copenhagen K, Denmark}

\author{Emily J. Safron}
\affiliation{Department of Physics and Astronomy, Louisiana State University, Baton Rouge, LA, USA}

\author{Zsofia Nagy}
\affiliation{Konkoly Observatory, Research Centre for Astronomy and Earth Sciences, H-1121 Budapest, Konkoly Thege \'ut 15--17, Hungary}



\begin{abstract}

We present observations toward HOPS 383, the first known outbursting Class 0 protostar located within the Orion molecular cloud using ALMA, VLA, and SMA. The SMA observations reveal envelope scale continuum and molecular line emission surrounding HOPS 383 at 0.85 mm, 1.1 mm, and 1.3 mm.The images show that HCO$^+$ and H$^{13}$CO$^+$ peaks on or near the continuum, while N$_2$H$^+$ is reduced at the same position. This reflects the underlying chemistry where CO evaporating close to the protostar destroys N$_2$H$^+$ while forming HCO$^+$. We also observe the molecular outflow traced by $^{12}$CO ($J = 2 \rightarrow 1$) and ($J = 3 \rightarrow 2$). A disk is resolved in the ALMA 0.87 mm dust continuum, orthogonal to the outflow direction, with an apparent radius of $\sim$62 AU. Radiative transfer modeling of the continuum gives disk masses of 0.02 M$_{\odot}$ when fit to the ALMA visibilities. The models including VLA 8 mm data indicate that the disk mass could be up to a factor of 10 larger due to lower dust opacity at longer wavelengths. The disk temperature and surface density profiles from the modeling, and an assumed protostar mass of 0.5  M$_{\odot}$ suggest that the Toomre $Q$ parameter $< 1$ before the outburst, making gravitational instability a viable mechanism to explain outbursts at an early age if the disk is sufficiently massive.
\end{abstract}



\section{Introduction} \label{sec:intro}

Protostellar disks around young stars found within the molecular clouds play a fundamental role in the formation and evolution of protostars (\citealt{Shu87}). These protostellar disks not only accrete mass onto their stars, driving the evolution of the protostars, but also provide the raw material for planet formation (\citealt{Testi14}). Starting with FU Orionis in 1936 (\citealt{Wachmann39}), several young stars have been observed in recent years to have luminosity bursts consistent with episodic accretion. However, it is still debated whether episodic accretion is an integral part of star formation, or if the star formation process is mostly smooth with only a few accretion bursts (\citealt{Dunham14}; \citealt{Audard14}). Determining the accretion history for stellar mass assembly is important because the physical and the chemical properties of the protostellar envelopes and their disks are affected by accretion bursts.

These episodic accretion bursts help to resolve the luminosity problem in Young Stellar Objects (YSOs). Simply stated, the luminosity problem results from observations of YSOs indicating that they are under-luminous compared to predictions from accretion models (\citealt{Kenyon90}; \citealt{Evans09}; \citealt{Kryukova12}). Mass dependent model of accretion or models with accretion bursts are able to solve the luminosity problem (\citealt{Dunham12}; \citealt{Fischer17}; \citealt{Offner11}), although in reality both process may occur. These accretion bursts dramatically increase the luminosity of the protostellar system which continues for decades to perhaps centuries (\citealt{Hartmann96}) and most of the stellar mass is accreted in these bursts. Most of these observed accretion bursts are from Class I/Class II, pre-main sequence YSOs. However, recent IR observations have extended these bursts into the protostellar phase (e.g., \citealt{Graham85}; \citealt{Reipurth04}; \citealt{Fischer12}; \citealt{Fischer19}; \citealt{Muzerolle13}). This is consistent with the idea that accretion bursts should be detectable throughout the protostellar phase (\citealt{Vorobyov06}).

Episodic accretion bursts have the ability to significantly alter the chemical composition of the protostellar envelope and the disk (e.g., \citealt{Visser12}; \citealt{Visser15}; \citealt{Jorgensen15}). These chemical effects primarily result from the change in temperature of the surrounding gas and dust due to the luminosity increase by $\sim$100$\times$. The increase in temperature can evaporate the CO (and other molecules) that were frozen on the grain mantles, releasing them back into the gas phase (\citealt{Lee07}; \citealt{Visser12}; \citealt{Visser15}; \citealt{Vorobyov13}). Since the freeze-out timescale of CO at protostellar envelope densities is longer than the duration of the bursts, the effects of the burst can be seen long after the actual burst has occurred (\citealt{Jorgensen15}; \citealt{Visser15}). The increase in luminosity can also alter the D/H ratio of water and other chemical species, triggering a series of changes in the molecular composition of the envelope (\citealt{Owen15}). Chemical signatures of outbursts have been detected in the past in low mass protostars. For example \citet{Jorgensen13, Jorgensen15} and \citet{Frimann17} found that nearly half of the Class 0/I protostars they observed had extended CO emission, corresponding to a transient rise in luminosity in the past 10$^{4}$ yr. Likewise, tracing CO and H$_{2}$O snowline, \cite{Hsieh18, Hsieh19} derived accretion burst interval of $\sim$2400 for Class 0 sources.

The youngest known YSO to have a detected outburst is the deeply embedded Class 0 protostar, HOPS 383 (\citealt{Safron15}). HOPS 383 is located in the Orion Molecular Cloud 3 (OMC3) region and was classified as a protostar in the \textit{Spitzer} (\citealt{Megeath12}) and \textit{Herschel Orion Protostar Survey} (\citealt{Furlan16}; \citealt{Fischer17}). It went through an outburst around the year 2006 which is most prominently seen at a wavelength of 24 $\mu$m. The luminosity of the protostar system increased from $\sim$0.2 L$_{\odot}$ to $\sim$7.5 L$_{\odot}$ in between 2004 and 2008. Furthermore, IR to sub-mm follow-up of HOPS 383 through 2010 shows no notable signs of decline in luminosity (\citealt{Safron15}). Given the duration of the outburst and the lack of decrease in luminosity, it is consistent with the characteristics of FU Orionis type outbursts, indicating that such outbursts are possible in the Class 0 phase.

Following the outburst, \citet{Madrid15} searched the Karl G. Jansky Very Large Array (VLA) archive for radio observations of HOPS 383 and were able to find only mild variations of flux from 1998 to 2014, suggesting that accretion and ejection enhancement do not follow each other, at least not within short periods of time. Comparing 850 $\mu$m James Clerk Maxwell Telescope (JCMT) Gould Belt Survey data with JCMT Transient Survey data obtained between 2012 and 2017, respectively, \citet{Mairs17} suggest that HOPS 383 could be a ``possible" variable candidate with a brightness decrease of 2.66\% $\pm$ 0.64\% yr$^{-1}$. More recently, \citet{Grosso20} suggest that the outburst of HOPS 383 peaked in 2008 and ended by 2017 September based on 4.6 $\mu$m observations of HOPS 383 with Wide-field Infrared Survey Explorer (WISE) and Near-Earth Object Wide-field Infrared Survey Explorer Reactivation (NEOWISE-R) between 2010 and 2019. On the other hand, most of the flux density at 4.6 $\mu$m is scattered light from the central protostar and disk, and can be affected by other factors than changes in the internal luminosity of the protostar.

The accretion burst in HOPS 383 provides a rare opportunity to observe the physical and chemical effects of an accretion burst on a protostellar envelope. Since the infalling envelopes are very dense in this phase, the continuum and molecular line emission of Class 0 objects are entangled with that of the disk (\citealt{Tobin18}). Thus, the properties of Class 0 disks such as their mass, radii, and temperature have not been determined for a broad sample. It is unclear if processes such as gravitational instability play an important factor at such a young stellar age. Constraining the cause, effects, and timescales of the outbursts would give us some predictive power to find other disks that might be capable of generating outbursts, or based on the chemical signatures around the portostar, predict if a system may have undergone outburst in the past.

In this paper, we report the  molecular line observations of $^{12}$CO, $^{13}$CO, C$^{18}$O, N$_2$H$^+$, HCO$^+$, and H$^{13}$CO$^+$ from the Submillimeter Array (SMA) in multiple configurations. We also present the continuum imaging from the Atacama Large Millimeter/submillimeter Array (ALMA) and the VLA and use these continuum observations to create a model of the disk and the envelope. The observations and the data reduction are described in Section \ref{sec:method}. The empirical results from the observations and the radio spectrum are presented in Section \ref{sec:results}. We detail the radiative transfer modeling of ALMA and VLA visibilities in Section \ref{sec:model}. The implications of our results are discussed in Section \ref{sec:discussion} and the conclusions are presented in Section \ref{sec:conclusion}.

\section{Observations and Data Reduction} \label{sec:method}

\subsection{SMA}
We used the SMA located on Mauna Kea, Hawaii in the Subcompact and the Extended configuration to observe HOPS 383. The Subcompact configuration was sensitive to the larger-scale emission at a resolution of 4$\arcsec$-7$\arcsec$ ($\sim$1700 - 3000 au). Combining with Extended configuration data provided an angular resolution of $\sim$1$\arcsec$ (420 au).

\subsubsection{Subcompact Observations}
The Subcompact observations were conducted on 27 October 2015 using 6 antennas and on 04 November 2015 using 8 antennas for a total time on source of 4.7 hr, 3.6 hr, and 7.5 hr for wavelengths of 850 $\mu$m, 1.1 mm and 1.3 mm, respectively. The flux, bandpass, and complex gain calibrations were carried out using Uranus, 3C279, and 0607-085, respectively. We used three different tunings centered at $\sim$225.5 GHz, $\sim$267.5 GHz, and at $\sim$351 GHz in the Subcompact configuration. The 225.5 GHz and 351 GHz observations were conducted simultaneously in dual-receiver, 2 GHz mode. This setup covered $^{12}$CO, $^{13}$CO, C$^{18}$O ($ J=2\rightarrow1$), HCO$^+$ ($ J=4\rightarrow3$), and H$^{13}$CO$^+$($ J=4\rightarrow3$). The 267.5 GHz tuning used a single-receiver, 2 GHz mode to observe the continuum and N$_2$H$^+$ ($ J=3\rightarrow2$), HCO$^+$ ($ J=3\rightarrow2$), and H$^{13}$CO$^+$ ($ J=3\rightarrow2$) lines. Our setup at 1.3 mm was also intended to observe the H$_2$CO ($ J=3_{0,3}\rightarrow2_{0,3}$), SiO ($ J=5\rightarrow4$), and SO ($ J_N=5_6\rightarrow4_5$) lines but we could not observe these lines due to problems with the correlator.

\subsubsection{Extended Observations}
The Extended observations were conducted on 20 November 2016 and on 27 November 2016 using 8 antennas for a total time on source of 5.9 hr and 2.8 hr for wavelengths of 1.1 mm and 1.3 mm, respectively. For the 20 November observation, Uranus was used as the flux calibrator and for the 27 November observation, both Uranus and Callisto were used as the flux calibrators. Bandpass and complex gain calibrations for both observations were carried out using 3C279 and 0607-085, respectively. We used two tunings centered at $\sim$225.5 GHz and at $\sim$267.5 GHz in the Extended configuration. We observed the same spectral setup as in the Subcompact configuration, but were able to also observe the H$_2$CO ($ J=3_{2,1}\rightarrow2_{2,1}$), SiO ($ J=5\rightarrow4$), and SO ($ J_N=5_6\rightarrow4_5$) lines.

\subsubsection{Data Reduction}
In both Subcompact and Extended Observations, all lines had a spectral resolution of 0.25 km s$^{-1}$, except $^{13}$CO and H$_2$CO, which had spectral resolution of 0.5 km s$^{-1}$ and $^{12}$CO, which had spectral resolution of 1 km s$^{-1}$. Calibration and editing of the SMA data were performed using the IDL based MIR software package. MIR was originally developed for the Owens Valley Radio Observatory and adapted by the SMA group. The estimated uncertainty of the absolute flux calibration is expected to be about 10-20\%. The imaging of both the continuum and the spectral line data were performed using the MIRIAD (Multichannel Image Reconstruction, Image Analysis and Display; \citealt{Sault95}) software package.

\subsection{ALMA}
HOPS 383 was observed on 2016 September 6 \& 2017 July 19 with ALMA located in the Llano de Chajnantor, Chile in band 7 (0.87 mm, Project code: 2015.1.00041.S) as part of the VLA/ALMA Nascent Disk and Multiplicity (VANDAM) Survey (\citealt{Tobin20}). The total on source time was $\sim$54 seconds. The baselines covered ranged from 15 m to 3697 m. 

The correlator setup for the 0.87 mm observations consisted of two of the basebands set to low spectral resolution 1.875 GHz continuum windows of 31.25 MHz channels. These basebands were centered at 333 GHz and 344 GHz. The third baseband was centered on $^{12}$CO ($J = 3 \rightarrow 2$) at 345.8 GHz with a 937.5 MHz spectral window (0.489 km s$^{-1}$ channels) and the last baseband was centered on $^{13}$CO ($J = 3 \rightarrow 2$) at 330.6 GHz with a 234.375 MHz (0.24 km~s$^{-1}$ channels) spectral window, although we were unable to detect it. The bandpass calibrator used during the two 2016 observation was J0510+1800, and J0522--3627 during the 2017 observation. J0510+1800, J0541--0541, and J0750+1231 were used as the absolute flux calibrators for the respective observations and J0541--0541 was used as the complex gain calibrator for all the observations. The estimated uncertainty of the absolute flux calibration is expected to be about 10\%.

To account for the variation of the quasar J0510+1800 used for flux calibration, the data were reduced manually by the Dutch Allegro ARC Node. All further processing was done using \textit{CASA} version 4.7.2 (\citealt{McMullin07}). The spectral line data were continuum subtracted using the task \textit{uvcontsub}, fitting the continuum to the line free channels in each spectral window. The data were imaged using the task \textit{clean} with a velocity resolution of 1 km s$^{-1}$ for $^{12}$CO, and 0.5 km s$^{-1}$ for $^{13}$CO. The signal-to-noise ratio of the continuum images was improved by performing self-calibration on the continuum and also applying these solutions to the spectral line data. The final continuum image was produced using Briggs weighting with a robust parameter of 0.5, providing a balance between sensitivity and resolution. This resulted in a synthesized beam of 0\farcs11 $\times$ 0\farcs10 and RMS noise of 0.28 mJy beam$^{-1}$.

\subsection{VLA}
HOPS 383 was observed with the VLA located in the plains of San Agustin, New Mexico, USA in the lower resolution ($\sim$0\farcs7) C-array on 2016 February 16 and in higher resolution ($\sim$0\farcs065) A-array on 2016 November 16. During each observation, 26 antennas were operating and the entire observation lasted 2.5 hours. The observations were conducted with the Ka-band receivers and three-bit correlator mode with one 4 GHz baseband centered at 36.9 GHz (8.1 mm) and the second 4 GHz baseband was centered at 29 GHz (1.05 cm). We used 3C48 (J0137+3309), 3C84 (J0319+4130), and J0541-0541 as the absolute flux, bandpass, and complex gain calibrators respectively. The estimated uncertainty of the absolute flux calibration is expected to be about 10\% for the VLA observations. To reduce the effect of rapid atmospheric phase variations in high-frequency observations, the observations were made in fast-switching mode (2.65 minute cycle times) and the total time on source was $\sim$64 minutes.

The data were reduced using the scripted version of the VLA pipeline in \textit{CASA} 4.4.0. The continuum was imaged using the \textit{clean} task in \textit{CASA} 4.5.1 using Natural weighting, multi-frequency synthesis with \textit{nterms=2} across both basebands. The final image has an RMS noise of 7.16 $\mu$Jy~beam$^{-1}$ and a synthesized beam of 0\farcs08$\times$0\farcs07 (32~AU~$\times$~28~AU).

\section{Results} \label{sec:results}
\subsection{Continuum}\label{subsec:continuum}

\begin{figure*}[ht!]
            \includegraphics[width=.49\textwidth]{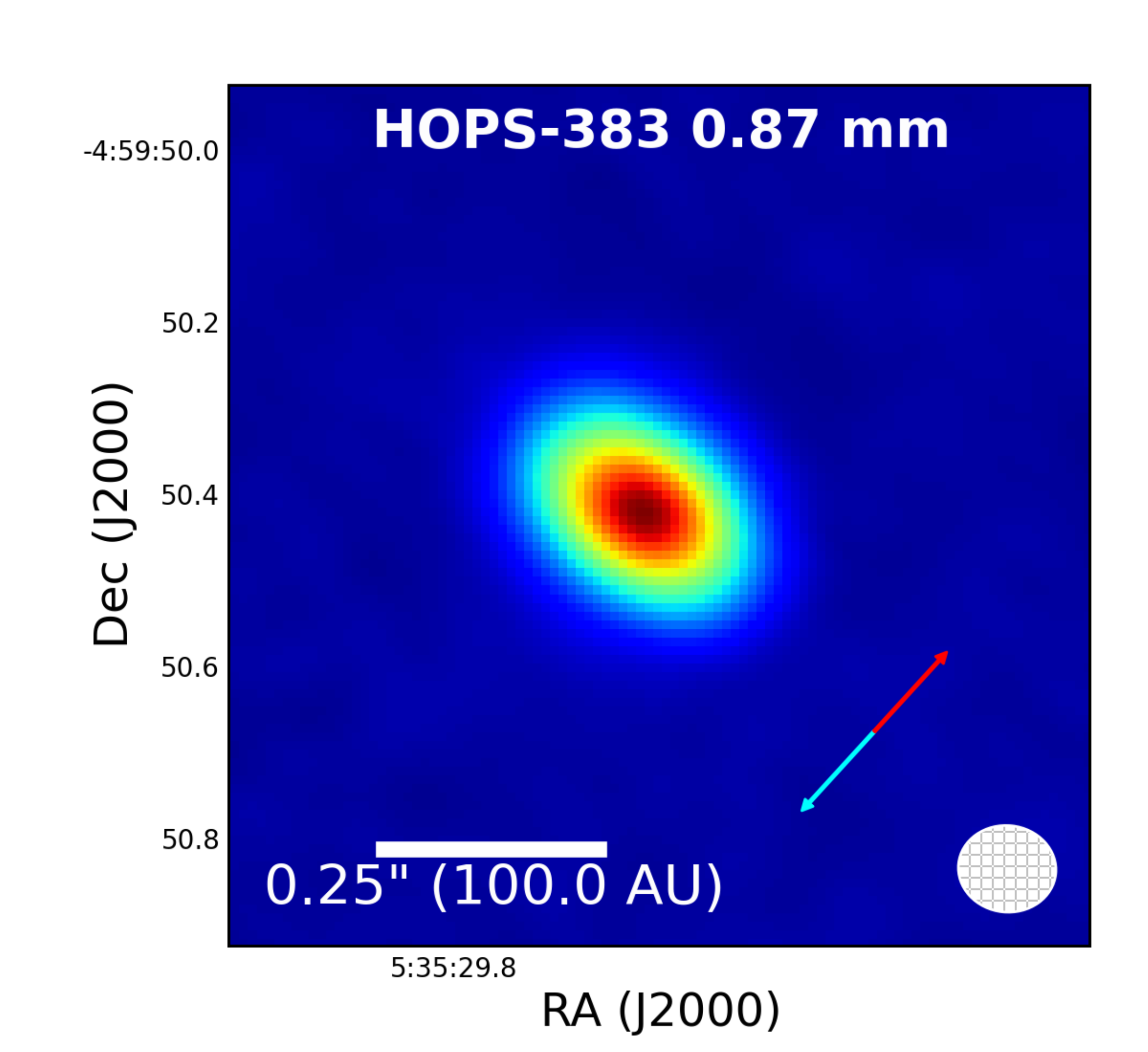}
            \includegraphics[width=.49\textwidth]{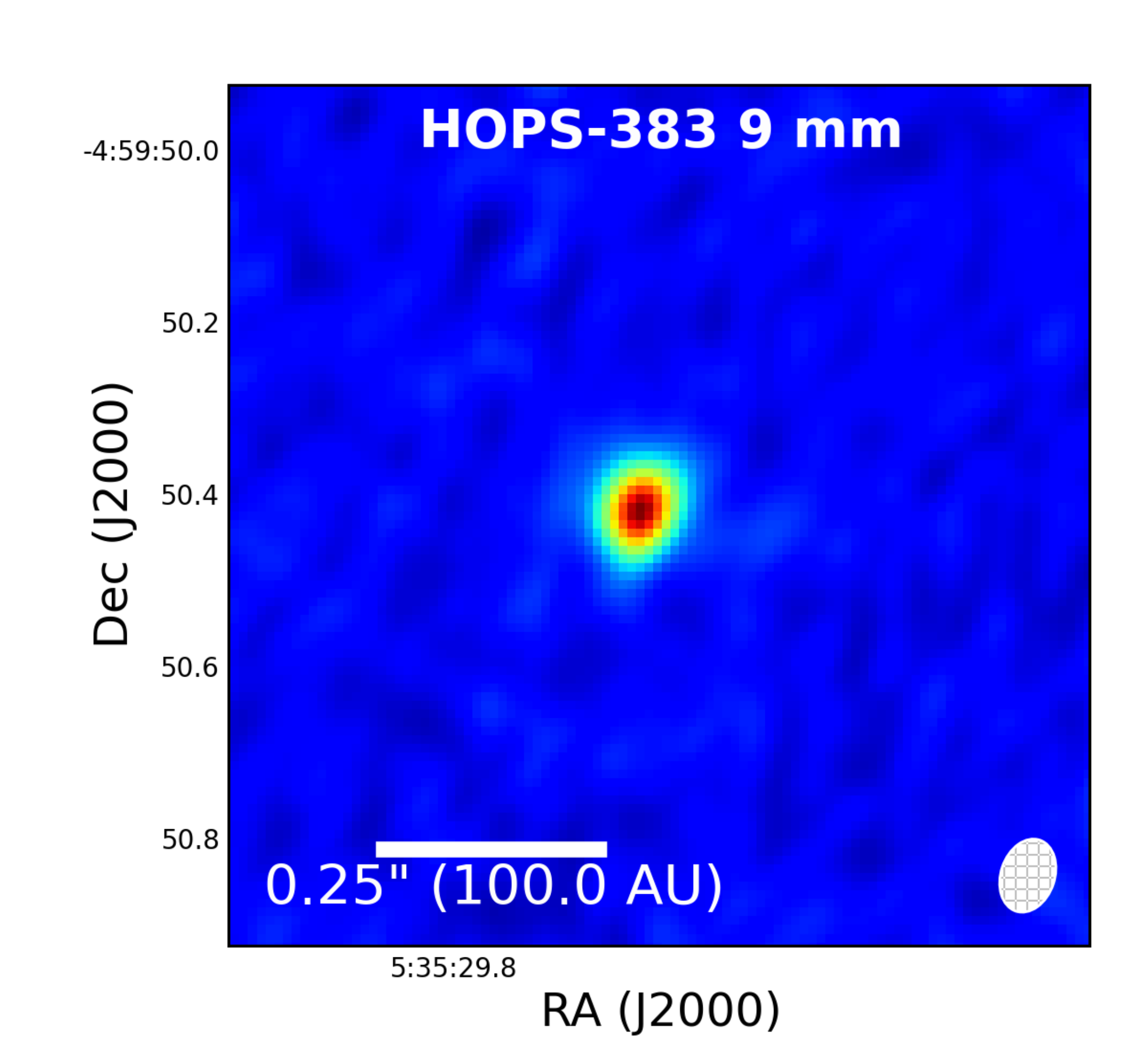}\linebreak\centering

\caption{HOPS 383 continuum images at 870 $\mu$m (\textit{left}) and 9 mm (\textit{right}) observed with ALMA, and the VLA respectively. The ALMA image clearly shows the source extended orthogonal to the outflow direction, marked by the red and blue arrows. The beam size in each figure is 0\farcs11 $\times$ 0\farcs10 and 0\farcs09 $\times$ 0\farcs06 at 0.87 mm and 9 mm respectively. \label{fig:continuum}}
\end{figure*}

We show the continuum images of HOPS 383 obtained from ALMA at 870 $\mu$m and the VLA at 9 mm in Figure \ref{fig:continuum}. The images show a very strong detection at these wavelengths, with the ALMA images showing that the dust continuum emission from HOPS 383 is resolved with its major axis oriented perpendicular to the outflow direction (Figure \ref{fig:12CO}). Thus, the resolved emission orthogonal to the outflow is consistent with the continuum emission tracing a resolved protostellar disk. The VLA image is more compact but is marginally resolved and has consistent position angle (PA) with the ALMA image. We observe a decrease in flux at the longest baselines from the VLA visibility profile (see Section \ref{sec:model}), indicating that we have resolved the continuum and peak intensity is less than the integrated flux density.

The disk radius is estimated to be $\sim$62 au from Gaussian fitting of the ALMA 870 $\mu$m continuum enables. This radius was originally reported by \citet{Tobin20} using the 2$\sigma$ of the Gaussian distribution radius derived from a single component Gaussian fit. The deconvolved FWHM source size is 0\farcs18$\times$0\farcs12 and a position angle of 49.7\degr; FWHM $\approx$ 2.355$\sigma$. The deconvolved size also enables us to estimate the inclination of HOPS-383 to be $\sim$48\degr\ calculated from arccos($\theta_{min}$/$\theta_{maj}$).

The integrated flux density of HOPS 383 at 870 $\mu$m is 157.8 $\pm$ 3.2 mJy with peak intensity value of 47 $\pm$ 0.7 mJy/beam. Likewise, the integrated flux density for the 8 mm wavelength data from the VLA is 411 $\pm$ 17 $\mu$Jy with peak intensity value of 290 $\pm$ 7 $\mu$Jy/beam. We also calculated the integrated flux density of HOPS 383 obtained from the SMA. In the Subcompact configuration, we obtained integrated flux density of 323.46 $\pm$ 12.7 mJy, 147.74 $\pm$ 12.2 mJy, and 136.63 $\pm$ 6.7 mJy at 0.85 mm, 1.1 mm, and 1.3 mm, respectively. Likewise, in the Extended configuration, we obtained integrated flux density of 128.81 $\pm$ 5.65 mJy and 75.72 $\pm$ 3.7 mJy at 1.1 mm and 1.3 mm, respectively. The SMA continuum is not as well resolved as the ALMA and the VLA. However, the lower flux density of SMA Extended configuration vs. Subcompact configuration indicates that the subcompact SMA fluxes contain a contribution from the envelope that is partially resolved out in the extended configuration data. The integrated flux densities from the SMA, ALMA, and the VLA observations are listed in Table \ref{table:flux}.

\begin{deluxetable*}{ccccccc}
\tablecaption{Integrated flux density of HOPS 383 obtained from various sub-millimeter to millimeter wavelength. \label{table:flux}}
\tablenum{1}
\tablehead{
\colhead{Wavelength} &
\colhead{Flux Density} &
\colhead{Peak Intensity} &
\colhead{RMS} &
\colhead{Instrument} &
\colhead{Configuration} &
\colhead{Beam Size}\\
\colhead{(mm)} & \colhead{(mJy)} & \colhead{(mJy bm$^{-1}$)} & \colhead{(mJy)} & \colhead{} & \colhead{} & \colhead{($\arcsec$)}
}
\startdata
0.85 & 323.46 $\pm$ 12.7 & 237.36 & 5.44 & SMA & Subcompact & 2\farcs92 $\times$ 2\farcs78 \\
0.87 & 157.75 $\pm$ 3.2 & 47.11 & 0.72 & ALMA & - & 0\farcs11 $\times$ 0\farcs10 \\
1.1 & 147.74 $\pm$ 12.2 & 76.73 & 3.4 & SMA & Subcompact & 6\farcs07 $\times$ 2\farcs93 \\
1.1 & 172.47 $\pm$ 8.9 & 131.72 & 3.93 & SMA & Subcompact w SWARM\tablenotemark{a} & 5\farcs94 $\times$ 2\farcs81 \\
1.1 & 128.81 $\pm$ 5.65 & 107.17 & 2.75 & SMA & Extended & 1\farcs18 $\times$ 0\farcs92 \\
1.3 & 136.63 $\pm$ 6.7 & 90.81 & 2.70 & SMA & Subcompact & 3\farcs77 $\times$ 3\farcs67 \\
1.3 & 75.72 $\pm$ 3.7 & 68.88 & 2.0 & SMA & Extended & 1\farcs46 $\times$ 1\farcs06 \\
8.1 & 0.500 $\pm$ 0.027 & 0.338 & 0.106 & VLA & AC combined & 0\farcs07 $\times$ 0\farcs06 \\
10.3 & 0.351 $\pm$ 0.024 & 0.238 & 0.094 & VLA & AC combined & 0\farcs10 $\times$ 0\farcs07 \\
9 & 0.411 $\pm$ 0.017 & 0.290 & 0.070 & VLA & AC combined & 0\farcs09 $\times$ 0\farcs06 \\
\enddata
\tablecomments{The uncertainties on the flux densities are statistical only and do not include $\sim$10\% absolute flux calibration uncertainty.}
\tablenotetext{a}{SMA Wideband Astronomical ROACH2 Machine}
\end{deluxetable*}

\subsection{Line Results}\label{subsec:line}

\begin{figure*}[ht!]
\includegraphics[width=0.5\textwidth]{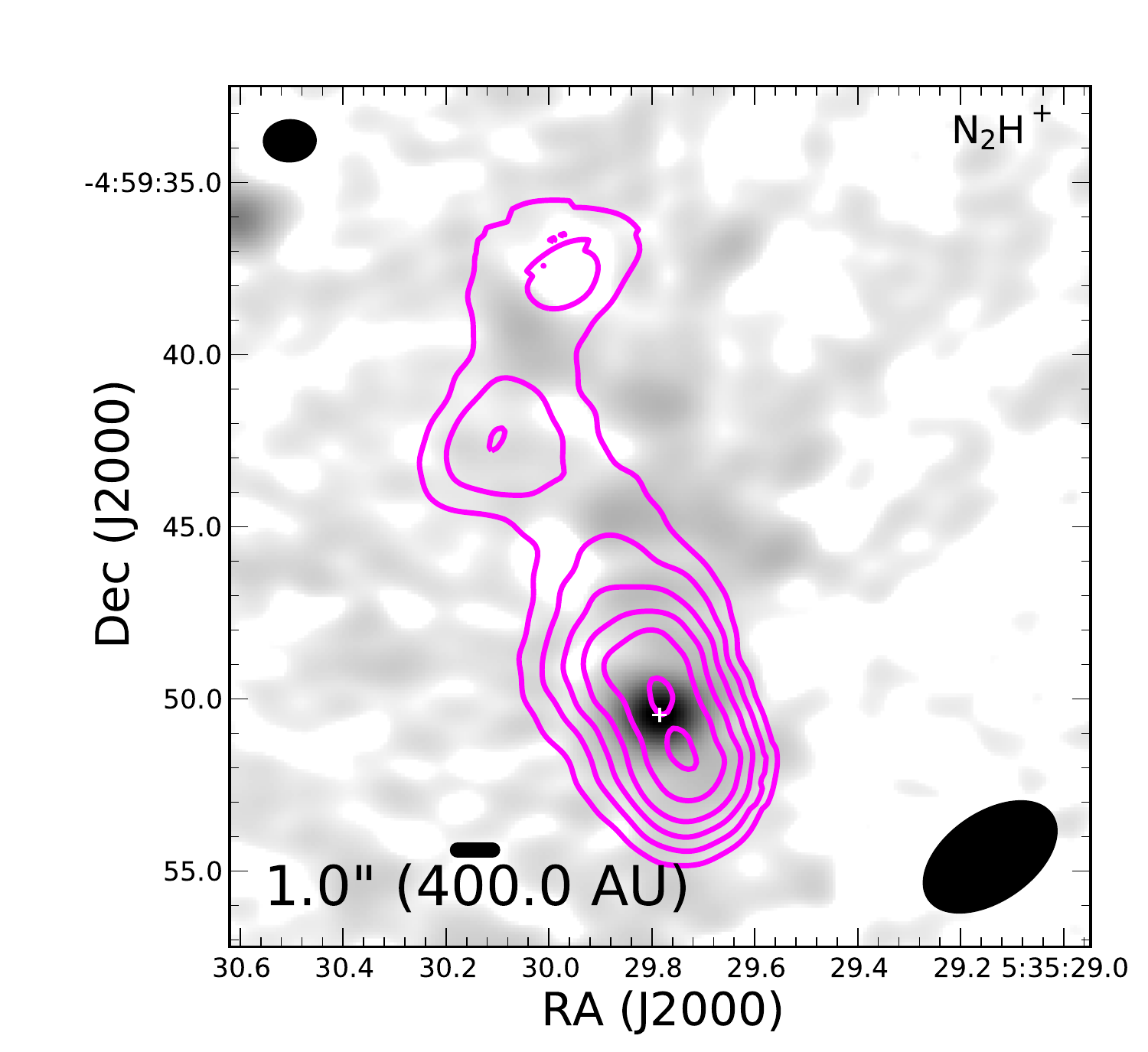}
\includegraphics[width=0.5\textwidth]{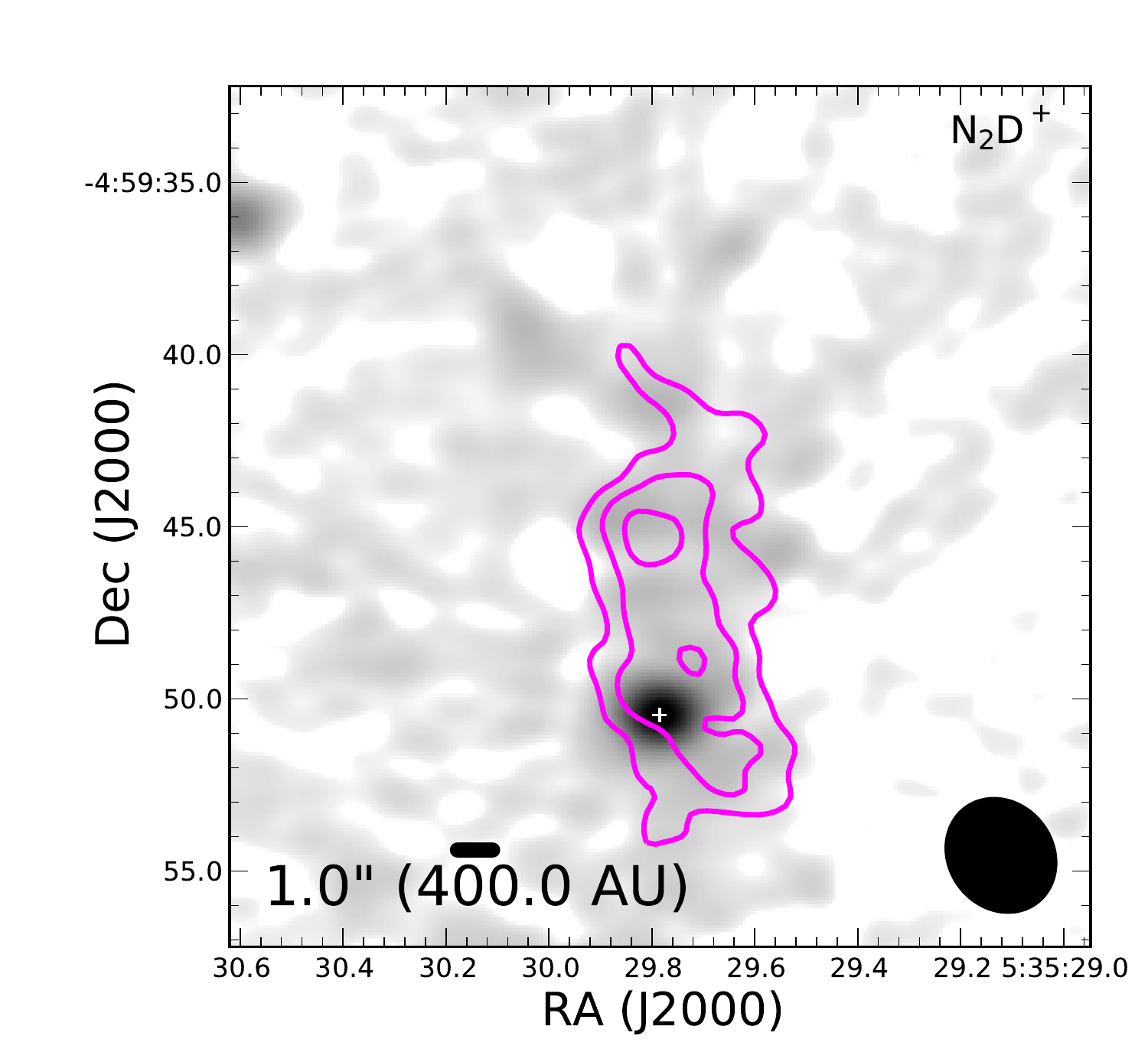}
\includegraphics[width=0.5\textwidth]{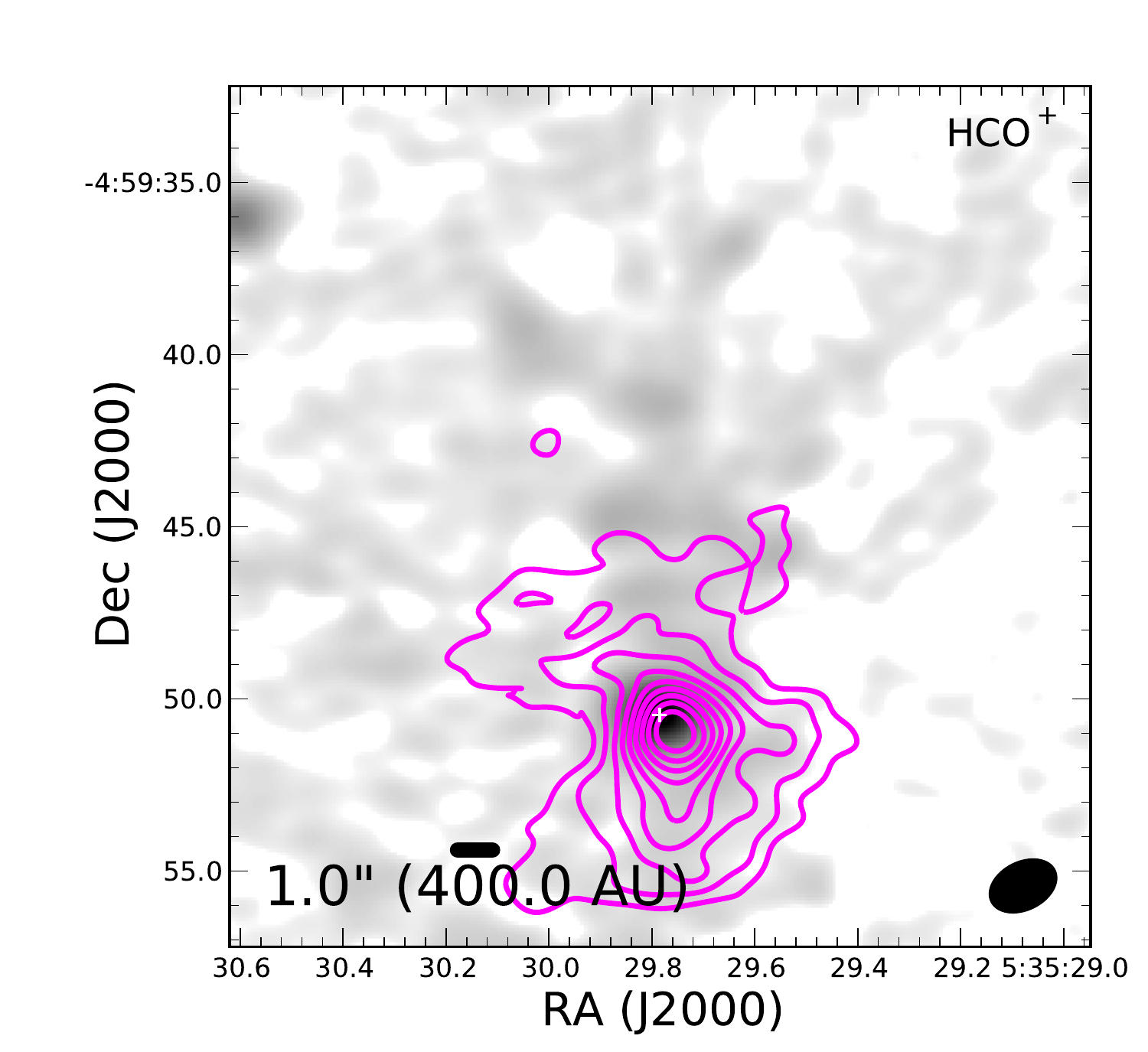}
\includegraphics[width=0.5\textwidth]{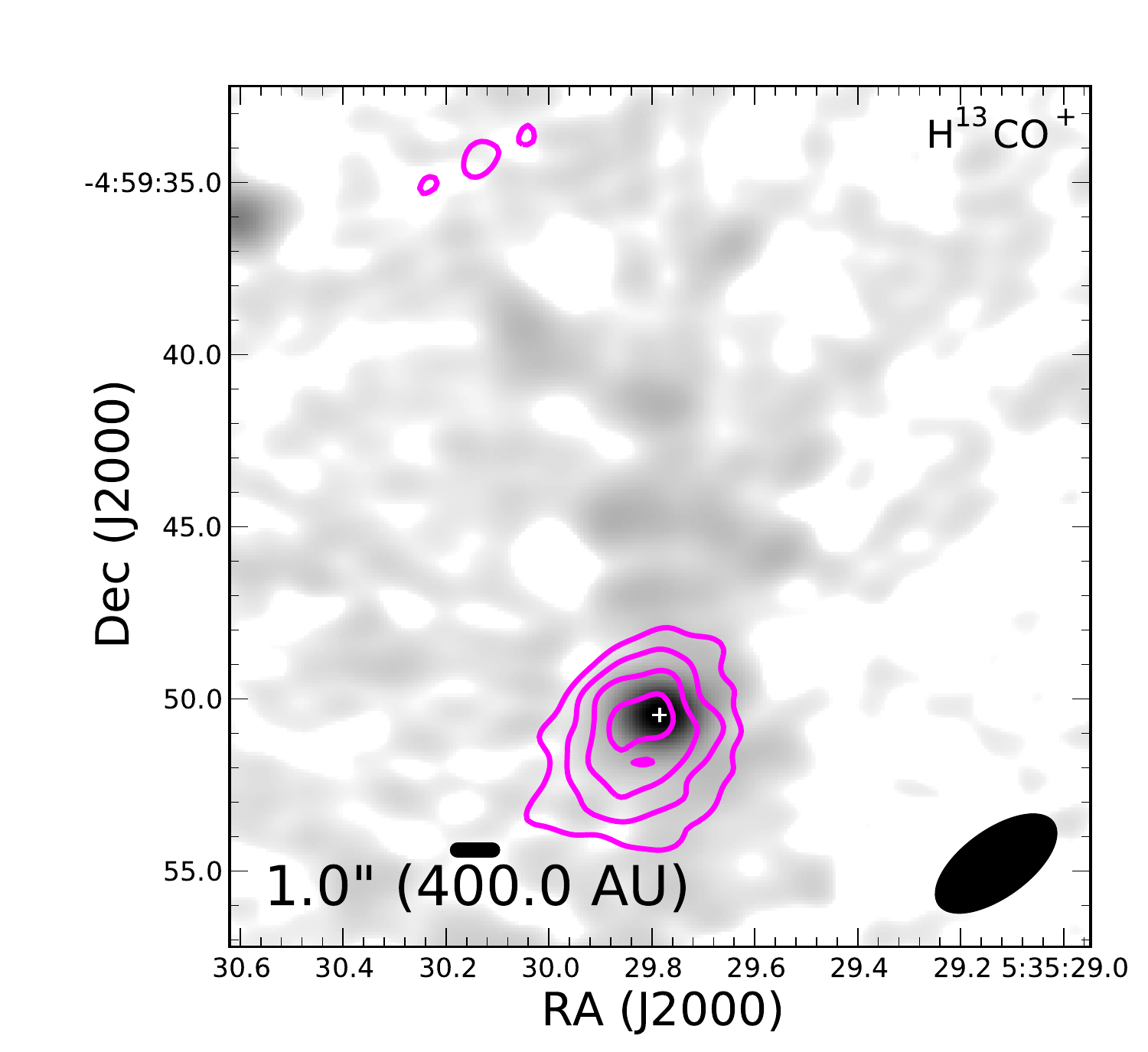}
\caption{Integrated intensity images from the combined Compact and Extended SMA observations of N$_2$H$^+$ ($ J=3\rightarrow2$), N$_2$D$^+$ ($ J=3\rightarrow2$) (\textit{up}), HCO$^+$ ($ J=3\rightarrow2$) and H$^{13}$CO$^+$ ($ J=3\rightarrow2$) (\textit{down}) overlaid on the continuum obtained from the combined Compact and Extended 1.3 mm SMA data. The contours start at 4$\sigma$ with steps of 2$\sigma$ for N$_2$H$^+$,HCO$^+$, and H$^{13}$CO$^+$ with $\sigma \approx$ 0.34, 0.15, and 0.095 Jy beam$^{-1}$ respectively and start at 3$\sigma$ with steps of 1$\sigma$ for N$_2$D$^+$ with $\sigma \approx$ 0.078 Jy beam$^{-1}$. The velocity ranges are 11.5 $\rightarrow$ 14.5 km/s, 12.75 $\rightarrow$ 13.5 km/s, 12.75 $\rightarrow$ 14.5 km/s, and 12.5 $\rightarrow$ 14.0 km/s for N$_2$H$^+$, N$_2$D$^+$, HCO$^+$, and H$^{13}$CO$^+$, respectively. The white cross represents the peak flux of the continuum used in the background. The beam size is 4$\farcs$3 $\times$ 2$\farcs$5 for N$_2$H$^+$, 3$\farcs$5 $\times$ 3$\farcs$0 for N$_2$D$^+$, 2$\farcs$0 $\times$ 1$\farcs$4 for HCO$^+$ and 4$\farcs$1 $\times$ 1$\farcs$9 for H$^{13}$CO$^+$. The background continuum beam size is 1$\farcs$5 $\times$ 1$\farcs$2 and is shown on the top left of the N$_2$H$^+$ image. \label{fig:line}}
\end{figure*}

The spectral setups of the SMA cover the N$_2$H$^+$, N$_2$D$^+$, HCO$^+$, and H$^{13}$CO$^+$ molecular lines, which were the key lines of interest. We show the integrated intensity (moment zero) maps from the combined Subcompact and Extended configuration in Figure \ref{fig:line}, overlaid on the 1.3 mm SMA continuum for these molecular lines. The peak signal-to-noise ratio of these transitions is $\sim$5-7 $\sigma$ in 0.25 kms$^{-1}$ channels. These molecules serve as complementary tracers that allow us to test the predictions of chemical models for outbursting systems by \citet{Visser15}.

N$_2$H$^+$ and N$_2$D$^+$ trace the cold dense regions ($n$ $\sim$10$^5$ cm$^{-3}$, $T \lesssim 20$ K) around the protostar and its envelope where CO is mostly frozen out (e.g. \citealt{Jorgensen04}; \citealt{Bergin07}; \citealt{Visser12}). We see N$_2$H$^+$ appearing marginally double-peaked about $\sim$250 au (north), and $\sim$500 au (south) straddling the protostar position in Figure \ref{fig:line}. HCO$^+$ mainly forms when CO reacts with H$_3^+$ and also as a by-product of gas-phase CO reacting with N$_2$H$^+$ (\citealt{Jorgensen04b}). HCO$^+$ should be present with a higher concentration inside R $\sim$500 au where the increase in luminosity converts the frozen out CO to gas-phase, converting N$_2$H$^+$ to HCO$^+$. In Figure \ref{fig:line}, we see that HCO$^+$ is peaked near the protostar position, corresponding to the warmer regions where we observe the decrease in N$_2$H$^+$. We also detect N$_2$D$^+$ and H$^{13}$CO$^+$ as shown in Figure \ref{fig:line} around HOPS 383 with similar spatial distributions to N$_{2}$H$^{+}$ and HCO$^{+}$, respectively. The surface brightness of N$_{2}$D$^{+}$ does not peak as strongly as that of N$_{2}$H$^{+}$, somewhat resembling the structures observed in N$_{2}$H$^{+}$ and N$_{2}$D$^{+}$ toward the Class 0 L1157-mm (\citealt{Tobin13_2}).

\begin{figure*}[ht!]
\includegraphics[width=1.0\textwidth]{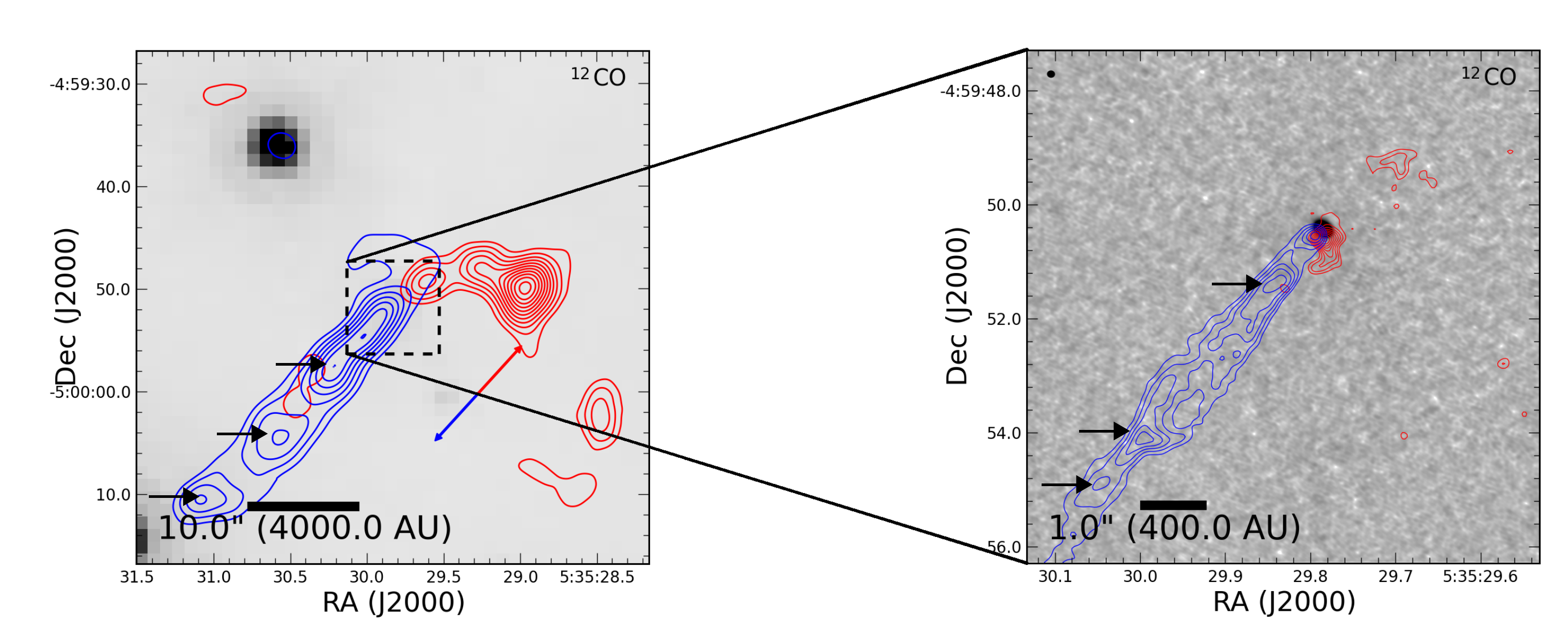}
\caption{HOPS 383 $^{12}$CO ($ J=3\rightarrow2$)  red- and blueshifted integrated intensity map tracing the outflow from the SMA 850 $\mu$m observations overlaid on the \textit{Spitzer}-IRAC 4.5 $\mu$m image (left). The velocity ranges are 0.0 $\rightarrow$ 10.0 km/s and 13.0 $\rightarrow$ 15.0 km/s for the respective blue- and redshifted emission. The right panel shows $^{12}$CO ($ J=3\rightarrow2$) emission observed with ALMA. The blueshifted contours start at 5$\sigma$ with steps of 2$\sigma$ with $\sigma \approx$ 0.15 Jy beam$^{-1}$. The redshifted contours start at 4$\sigma$ with steps of 1$\sigma$ with $\sigma \approx$ 0.045 Jy beam$^{-1}$. The black arrows show the possible knots that appear on both the SMA and the ALMA observations.} \label{fig:12CO}
\end{figure*}

The SMA spectral setup also clearly detects the molecular outflow in $^{12}$CO ($J=3\rightarrow2$). The integrated intensity maps of the blue- and redshifted outflow from HOPS 383 are shown in Figure \ref{fig:12CO}. Based on the orientation of the outflow, we determine the position angle (PA) to be about 137.5$^{\circ}$. \citet{Feddersen20} found the PA of HOPS 383 to be 128 $\pm$ 0.5$^{\circ}$ based on observations from the Combined Array for Research in Millimeter-wave Astronomy-Nobeyama Radio Observatory (CARMA-NRO) Orion survey However, their observations had lower resolutions compared to ours. Hence, the small difference is reasonable.

\begin{figure*}[ht!]
\includegraphics[width=0.5\textwidth]{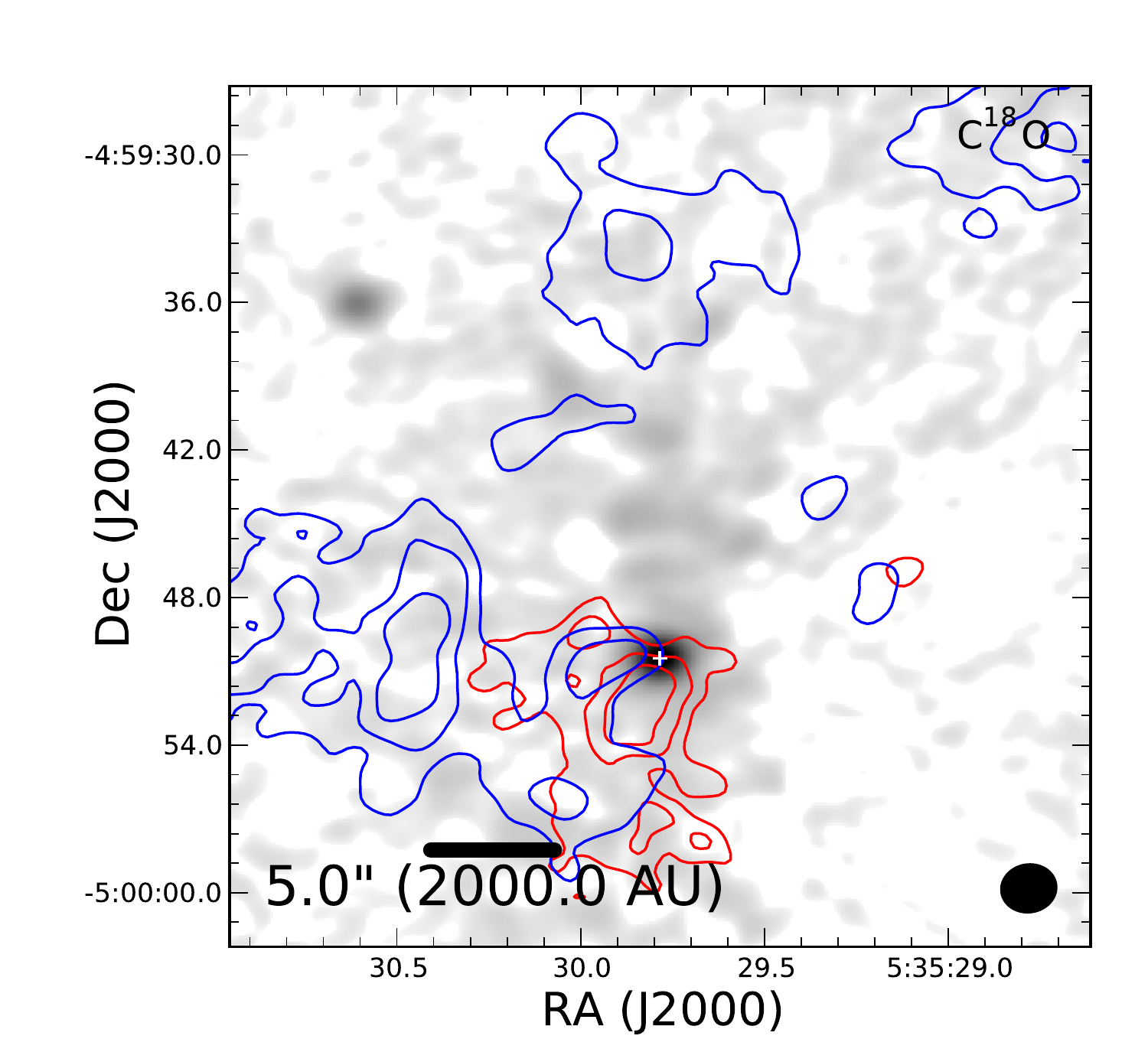}
\includegraphics[width=0.5\textwidth]{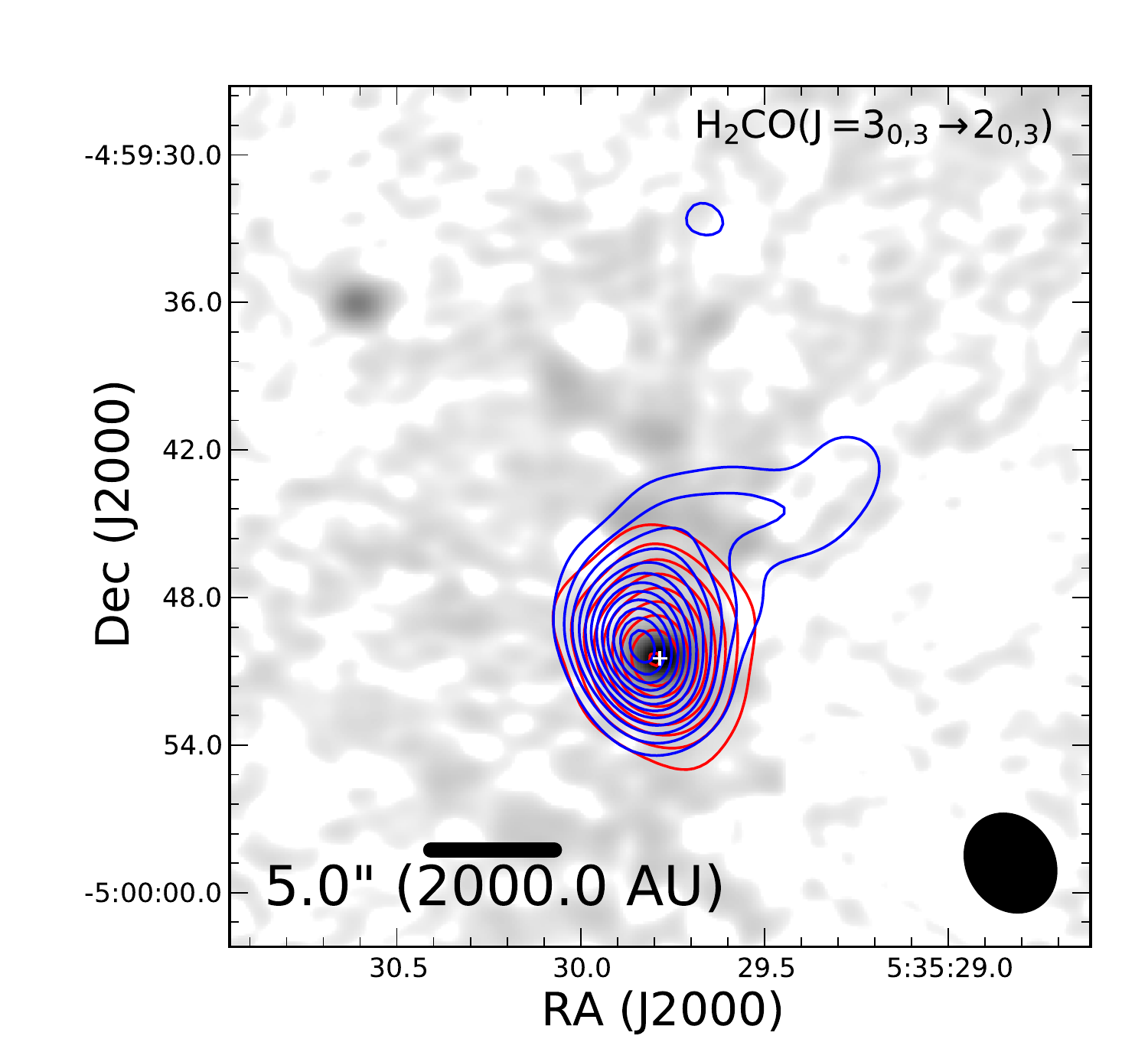}
\caption{HOPS 383 red- and blue-shifted integrated intensity map tracing the
emission from C$^{18}$O (left) and H$_2$CO (right) from the SMA observations overlaid on
the continuum obtained from the combined Compact and Extended 1.3 mm SMA
data. The C$^{18}$O contours start at 4$\sigma$ with steps of 2$\sigma$ with $\sigma \approx$ 0.06 Jy beam$^{-1}$ and $\sigma \approx$ 0.10 Jy beam$^{-1}$ for the red- and blue-shifted emission, respectively. The H$_2$CO contours start at 4$\sigma$ with steps of 2$\sigma$ with $\sigma \approx$ 0.05 Jy beam$^{-1}$ and $\sigma \approx$ 0.03 Jy beam$^{-1}$ for the red- and blue-shifted emission, respectively. The velocity ranges are 12.5 $\rightarrow$ 14.0 km/s and 12.75 $\rightarrow$ 14.0 km/s for C$^{18}$O and H$_2$CO respectively. The white cross represents the peak flux of the continuum used in the background. The beam size is 2$\farcs$2 $\times$ 1$\farcs$9 for C$^{18}$O and 4$\farcs$2 $\times$ 3$\farcs$5 for H$_2$CO. The background continuum beam size is 1$\farcs$5 $\times$ 1$\farcs$2, the same as on Figure \ref{fig:line}. \label{fig:c18o}}
\end{figure*}

We also detected the C$^{18}$O ($J =$ 2 $\rightarrow$ 1) and H$_2$CO ($ J=3_{0,3}\rightarrow2_{0,3}$) emission in our spectral setup. Figure \ref{fig:c18o} shows the integrated intensity map of the blue- and red-shifted emission from the combined Extended and Subcompact configuration for the C$^{18}$O and the Subcompact configuration for the H$_2$CO overlaid on the 1.3 mm SMA continuum. \citet{Jorgensen15} found that an accretion burst increases the sublimation radius of CO around the protostar. However, our C$^{18}$O data cube is contaminated by a foreground cloud that is also emitting at a similar velocity range, which mostly affected the blue-shifted emission. Therefore, the C$^{18}$O emission must be used with caution. Comparing with Figure \ref{fig:12CO}, unlike $^{12}$CO, C$^{18}$O is not significantly affected by the outflow from HOPS 383. This is mainly because C$^{18}$O is mostly optically thin and highly sensitive to the inner 1000 au radii near where CO is evaporated from the dust grains (\citealt{Jorgensen15}). H$_2$CO is peaked near the protostar from a smaller radius than that of CO showing that it requires a warmer temperature to sublimate. Experiments by \citet{Noble12} and \citet{Fedoseev15} found binding energy of H$_2$CO $\sim$3300 K giving freeze-out temperature of $\sim$70 K. 

\begin{figure}[ht!]
\includegraphics[width=1.0\linewidth]{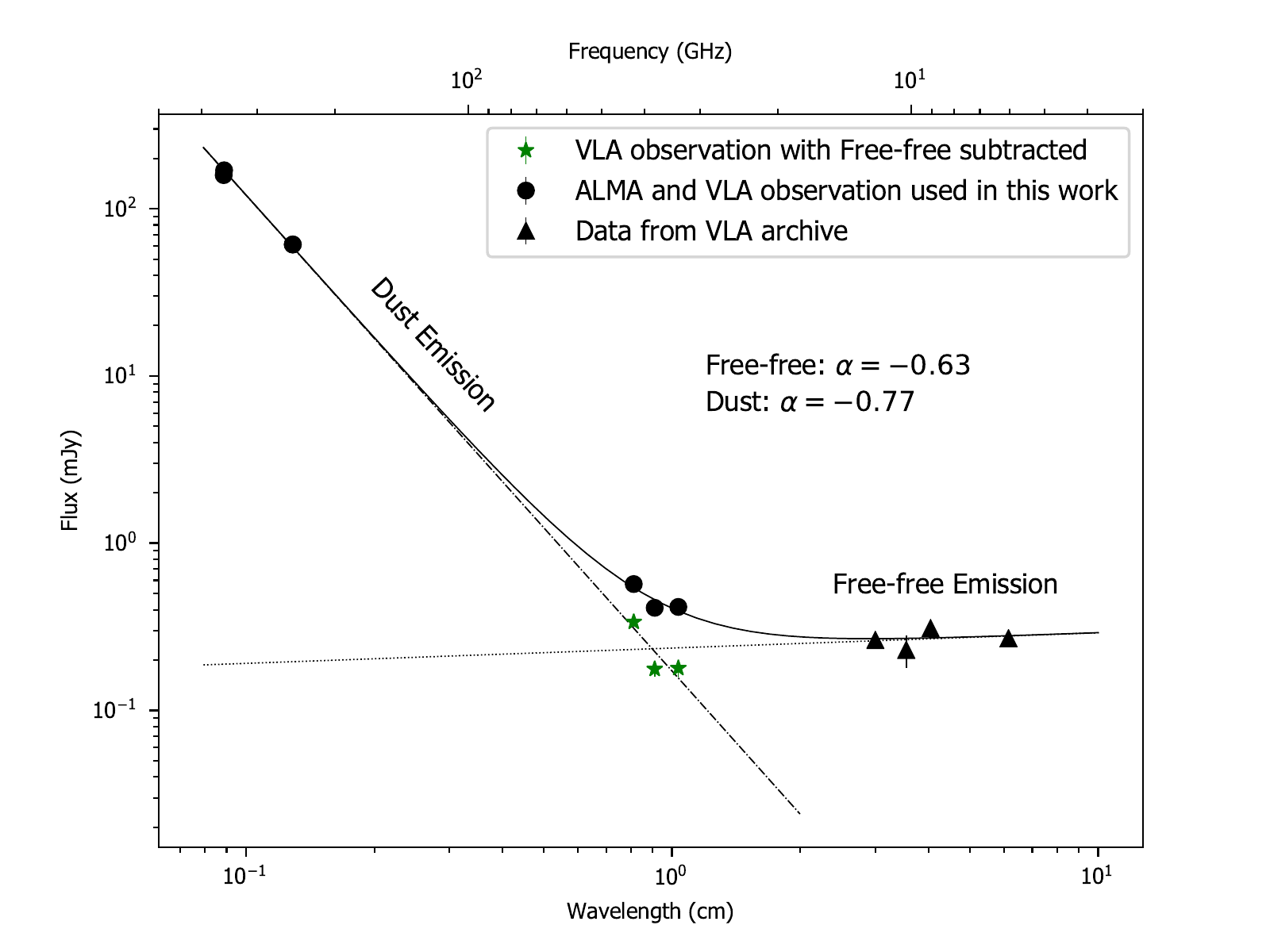}
\caption{The observed millimeter to centimeter spectrum of HOPS 383. The black circles are the data from the ALMA and the VLA used in this work. The black triangles are data taken from the VLA archive as shown in Table 1 of \citet{Madrid15}. We exclude the points with only upper limits and flux data before the outburst. We also only include the average of all the flux data for the 10 GHz archival data. The green stars represent the VLA fluxes used in this work with the extrapolated free-free emission subtracted.\label{fig:SED}}
\end{figure}

\subsubsection{Knotty structures in the $^{12}$CO outflow}\label{subsubsec:knots}
Both the low- and high-resolution outflow maps of HOPS 383 show some evidence of clump $^{12}$CO emission. We denote these structures with black arrows in Figure \ref{fig:12CO}. Such clumpy emission or knots may be related to time-variable accretion onto the protostar (\citealt{Plunkett15}; \citealt{Reipurth97}). Hence, the observation of knots in the outflow may help us constrain the number and timescale of previous accretion bursts (\citealt{Hsieh16}). \citet{Plunkett15} observed many knots in the outflow from the Class 0 protostar CARMA 7, having dynamical timescales of $\sim$100 yr to 6000 yr. \citet{Vorobyov18} compared the dynamical timescales of the outflow knots to the accretion burst timescales from their numerical hydrodynamical simulations, finding evidence for accretion bursts in gravitationally unstable disks happening on similar timescales to outflow knot ejections. From the number of bursts detected in Orion, \citet{Fischer19} inferred a burst rate of 1 per 1000 year for a protostar.

Since the observations used in this paper are only about 10 yr after the burst, any knots produced from the burst have not had enough time to be resolved with the resolution of the observation. The knot closest to the protostar is about 1$\farcs$35 away. We can roughly calculate the dynamic timescale of the closest knot using the spectral cube from ALMA and the formula $\tau_{dyn} = D/V_{flow} (cos~i/sin~i)$ (\citealt{Plunkett15}). Here, $D$ is the distance between the knot and the protostar, $V_{flow}$ is the line-of-sight velocity of the knot, and $i$ is the inclination of the outflow to the line of sight. We get $\tau_{dyn} \approx$ 215 yr, with $V_{flow} \approx 10$ km/s for the inclination derived at Section \ref{subsec:continuum}. Thus, the most recent clump in the outflow was ejected too long ago to be associated with the most recent accretion burst. However, if there is a one-to-one relationship between accretion bursts and outflow knots, there could have been past outbursts in HOPS 383. 

We can also roughly estimate when the other knots in the SMA plots were ejected. Assuming the knots travel at a constant speed after being ejected, they are traveling at a speed of $\sim$1 arcsec every 160 yr. The knots on the SMA are roughly 9 arcsec apart from one another, which translates to roughly 1400 yr between the knots. This aligns well with the rate of bursts inferred by \citet{Fischer19}. 

\subsection{Millimeter to Centimeter Spectrum}\label{subsec:sed}
Using the data from ALMA and the VLA, we extend the radio spectrum of HOPS 383 from millimeter to centimeter wavelengths. Figure \ref{fig:SED} shows the spectrum of the protostar at the millimeter wavelengths along with a few centimeter wavelengths data from the VLA archive reported by \citet{Madrid15}. From the archival data, we averaged the 2014 data and discarded the upper-limits (see Table 1, \citealt{Madrid15}). HOPS 383 is not at the phase center of these archival data but lies within the field of view.  

Figure \ref{fig:SED} shows that the spectrum at millimeter wavelength has a steeply declining slope with respect to wavelength indicative of the emission being dominated by dust emission while centimeter wavelengths have a flat slope indicative of the emission being dominated by free-free emission. The 8 mm to 1 cm data falls in the transition from dust to free-free emission. We use the \textit{curve\_fit} function from Python's \textit{scipy} library to fit the free-free emission power law to the longer wavelength data and subtract the extrapolated contribution from the free-free emission to the flux densities on the 8 mm to 1 cm data. We then fit another power law for the dust emission from 0.87 mm to 1 cm. The free-free contribution to the 8 mm, 9 mm, and 1 cm data is found to be $\sim$0.232 mJy, $\sim$0.234 mJy, and $\sim$0.237 mJy, respectively. From the spectrum, we get the power law as $\text{F}_{\lambda, \text{dust}} = (0.17 \pm 0.08)\lambda^{-2.85 \pm 0.05}$ mJy and $\text{F}_{\lambda , \text{ff}} = (0.24 \pm 0.09)(\lambda/10.0 \text{mm})^{0.09 \pm 0.27}$ mJy.

\subsection{Disk Mass}\label{subsec:mass}

The resolved dust continuum emission from the VLA and ALMA can be used to estimate the mass of the total disk structure surrounding HOPS 383. Assuming isothermal dust emission, well-mixed gas and dust, and optically thin emission, the disk mass is given by,
\begin{equation}
M_{\text{dust}}=\frac{D^{2}F_{\lambda}}{\kappa_{\lambda}B_{\lambda}(T_{\text{dust}})},
\end{equation}
where D is the distance to the HOPS 383 ($\sim$400 pc), $T_{\text{dust}}$ is the dust temperature, $F_{\lambda}$ is the flux density observed at $\lambda$, $\kappa_{\lambda}$ is the dust opacity, and $B_{\lambda}$ is the Planck function at the wavelength, $\lambda$. We take $\kappa_{1.3 \text{ mm}} = 0.899$ cm$^2$g$^{-1}$ for thin ice mantles from dust opacity models of \citet{Ossenkopf94} and scale the opacities to the wavelength of ALMA and VLA using the equation,
\begin{equation}\label{eqn:opacity}
\kappa_{\lambda}=\kappa_{1.3\text{ mm}}\times\Bigg(\frac{1.3\text{ mm}}{\lambda}\Bigg)^{\beta},
\end{equation}
where $\beta$ is the dust opacity spectral index. For a typical protostar with low luminosity ($\sim$1 - 3 L$_{\odot}$), T$_{\text{dust}}$ is assumed to be constant at 30 K, compatible with protostar disk estimates on $\sim$100 au scales (\citealt{whitney03}). We can thus scale the average T$_{dust}$ for HOPS 383 using (L/L$_{\odot}$)$^{0.25}$, as expected from thermal equilibrium. This gives us an average temperature of $\sim$50 K for L$_{bol}$ = 7.8 L$_{\odot}$, which we use as a high temperature estimate for HOPS 383. Assuming canonical ISM dust to gas mass ratio of 1:100 (\citealt{Bohlin78}), we multiply the resulting dust mass by 100 to get the total disk mass.

The continuum flux density from ALMA at 870 $\mu$m has a flux density of 158 $\pm$ 3.2 mJy. Assuming $\beta\sim1.78$, which is appropriate of young, embedded cores, we obtain disk mass of 0.082 M$_{\odot}$ and 0.043 M$_{\odot}$ for T$_{\text{dust}}$ = 30 K and T$_{\text{dust}}$ = 50 K, respectively. The dust emission may not be optically thin at 0.87 mm for all radii, thus, the disk masses are likely lower limits for their respective temperatures. Based on the study of 328 protostars in Orion, \citet{Tobin20} found that the mean disk mass of Class 0 protostars was 0.0078 M$_{\odot}$. HOPS 383 appears to be more massive than the mean Class 0 disk based on ALMA derived flux density.

The continuum flux density from the VLA at 9 mm has a flux density of 411 $\pm$ 1.7 $\mu$Jy. Assuming $\beta\sim1$ here, which yields more consistent results between 1.3 mm and 1 cm (\citealt{Tobin16}; \citealt{Tychoniec18}), we obtain total disk mass of 0.25 M$_{\odot}$ and 0.15 M$_{\odot}$ for T$_{\text{dust}}$ = 30 K and T$_{\text{dust}}$ = 50 K, respectively. Subtracting the contribution of the free-free emission at 9 mm (0.234 mJy), we get total disk mass of 0.11 M$_{\odot}$ and 0.063 M$_{\odot}$ for T$_{\text{dust}}$ = 30 K and T$_{\text{dust}}$ = 50 K, respectively. Comparing with the results from \citet{Tobin20}, we again find that the HOPS 383 is more massive compared to the average mass of Class 0 disks. The dust emission at 9 mm is likely more optically thin than the dust emission at 870 $\mu$m, which is why it is tracing higher dust mass. This may explain why we measure larger masses at 9 mm, assuming that our adopted dust opacities at both wavelengths are correct.

\section{Continuum Modeling} \label{sec:model}
To understand the cause of the outburst and study the features of HOPS 383 in detail, we use Monte Carlo radiative transfer modeling to create synthetic models of the protostellar systems that match the visibilities from the VLA and/or ALMA and the SED of HOPS 383. We followed the modeling procedure of \citet{Sheehan14} and \citet{Sheehan17}. The models have a protostar at the center, surrounded by a circumstellar disk and an envelope with an outflow cavity. The different components and the free parameters of the models, along with a brief outline of the modeling technique are described below.

\subsection{Protostar}\label{subsec:protostar}
Our models include a central protostar with a temperature of 4000 K, representative of a general low mass star in the pre-main sequence phase. Since the T$_{\text{eff}}$ of Class 0 YSOs such as HOPS 383 have not been classified, this is a reasonable estimate to use for the models as nearly all emission in its SED is reprocessed; a small fraction of stellar light might be observed as scattered light in the outflow cavity. The luminosity, $L_{*}$, which includes stellar luminosity and accretion luminosity, is left as a free parameter in all our models. 

\subsection{Protostellar Disk} \label{subsec:diskmodel}
The protostellar disk model uses the standard density profile of a flared exponentially-tapered accretion disk (e.g. \citealt{lynden74}),
\begin{equation}
\rho = \rho_0 \Bigg(\frac{R}{R_0}\Bigg)^{-\alpha} exp\Bigg(-\frac{1}{2}\Bigg[\frac{z}{h(R)}\Bigg]^2\Bigg)exp\Bigg(-\Bigg[\frac{R}{R_c}\Bigg]^{2 - \gamma} \Bigg),
\end{equation}
\begin{equation}
h(R)=h_0\Bigg(\frac{R}{1 \text{ au}}\Bigg)^\beta.
\end{equation}
Here, $R$ and $z$ are the usual variables in cylindrical coordinates. $\rho_0$ is the density of the midplane at a radius of $R_{0}$ = 1 AU. Similarly, $h(R)$ is the disk scale height, and $h_0$ is the scale height at 1 AU. We truncate the inner disk at a specific inner radius, $R_{in}$, which is left as a free paramater. 

We define the disk surface density profile as,
\begin{equation}
\Sigma=\Sigma_0\Bigg(\frac{R}{R_0}\Bigg)^{-\gamma}exp\Bigg(-\Bigg[\frac{R}{R_c}\Bigg]^{2 - \gamma} \Bigg).
\end{equation}
Here, $\gamma$ is the surface density power-law exponent given by \(\gamma=\alpha-\beta\), and $\beta$ is scale height power law. For the outer radius of the disk, $R_{disk}$, we run both models with $R_{disk}$ truncated at a specific outer radius and with $R_{disk}$ exponentially tapered. In exponentially tapered models, $R_{disk}$ signifies where the exponential taper starts. We model two different types of disks as it is unclear which type of disk better reflects the underlying protostellar disk structure. Similarly, in the models we treat the disk mass, $M_{disk}$, $h_0$, $\gamma$, $\beta$ as free parameters as well.

\subsection{Envelope}\label{subsec:envelope}
A Class 0 protostar such as HOPS 383 is embedded in an envelope extending beyond the protostellar disk. We model the envelope using the solution for the density profile of a rotating collapsing envelope found by \citet{Ulrich76},
\begin{equation}
\rho=\frac{\dot{M}}{4\pi (GM_*r^3)^\frac{1}{2}}\Bigg(1+\frac{\mu}{\mu_0}\Bigg)^{-\frac{1}{2}}\Bigg(\frac{\mu}{\mu_0}+2\mu_0^2\frac{R_c}{r}\Bigg)^{-1}.
\end{equation}
Here $M_*$ is the mass of the protostar only used for setting density structure and is not constrained by fitting, $\dot{M}$ is the mass infall rate from envelope to disk and $R_c$ is the centrifugal radius inside which the envelope material can be rotationally supported (\citealt{Ulrich76}; \citealt{Terebey84}). In our model $R_c=R_{disk}$, $\mu=cos\theta$, and $r$ and $\theta$ are the usual variables in spherical coordinates. $\mu_0 = cos\theta_0$ is the initial angle of the infalling material (\citealt{Ulrich76}). The envelope in our models is truncated at the same inner radius, $R_{in}$, as the disk and an outer radius, $R_{env}$. The envelope mass $M_{env}$ and $R_{env}$ are left as free parameters.

Our envelope models have an outflow cavity, whose location is determined by,

\begin{equation}
    z > 1\text{ au} + r^{\xi}
\end{equation}
where $r$ is in the midplane (\citealt{Eisner12}). Inside the outflow cavity, the envelope density is reduced by the scaling factor for the density, $f_{\text{cav}}$. Likewise, $\xi$ is related to the outflow cavity opening angle as,

\begin{equation}
    \text{Opening angle} =2 \tan^{-1}\Big({\huge R}_{\text{env}} ^{(1/\xi -1)}\Big)
\end{equation}
and is left as a free parameter in all our models along with $f_{\text{cav}}$. We also leave the viewing angle parameters, inclination of the system ($i$) and position angle (p.a.) as free parameters in our models.

\subsection{Dust}\label{subsec:dust}
Our disk models use the dust opacities from \citet{Woitke16}. The dust grain models consist of 60\% silicate, 15\% amorphous carbon, and 25\% porosity. Our models follow a power-law grain size distribution with $n \propto a^{-p}$ with $p$ = 2.5 to 4.5, based on $p$ = 3.5 given by \citet{Mathis77}, where $a$ is the dust grain size. The minimum grain size is assumed to be 0.005 $\mu$m for all opacities. We use maximum dust grain size of 1 $\mu$m for the dust in the envelope. However, in the disk, the maximum size of the dust is left as a free parameter.

\subsection{Radiative Transfer Code}\label{subsec:rte}

We use RADMC-3D (\citealt{Dullemond12}) for the continuum modeling and to calculate the temperature throughout the density structure of the disk and envelope. The results are then used to produce synthetic observations of HOPS 383 that are simultaneously fit to either the 0.87 mm visibilities or both the 0.87 mm and 8 mm visibilities plus the spectral energy distribution (SED) using Markov Chain Monte Carlo (MCMC) fitting with the \textit{emcee} code (\citealt{Foreman13}). Once we have the temperature and the density, synthetic SEDs and millimeter images are produced using ray tracing in RADMC-3D. Synthetic visibilities are then created by Fourier-transforming the synthetic millimeter images. 

The MCMC walkers use Bayesian statistics to move through parameter space and reach a steady-state to provide estimates on the best-fit value for each of the free parameters. They aim to maximize the model likelihood, or equivalently, minimize the $\chi^2$ of the model in log-space. We use the weighted sum of the $\chi^2$ values of visibilities and the SED,

\begin{equation}
    X^2 = w_{\text{vis}}\chi^2_{\text{vis}} + w_{\text{SED}}\chi^2_{\text{SED}}
\end{equation}
where $w$ is the weight of each data set. Our fits use $w_{\text{vis}} = 0.5$ and $w_{\text{SED}} = 1$ for both the ALMA only and the ALMA+VLA models, which we empirically determined to fit both data sets simultaneously. The walkers converge towards the regions of parameter space which fit the data well. The best-fit value is considered as the median position of the walkers once it has reached a steady-state and the parameter changes only marginally over a large number of steps, i.e. the best fit values of the parameters remain more or less constant for $\sim$ 150 - 200 steps. The 1$\sigma$ value around the median containing $\sim$68\% of the walkers is considered as the uncertainty on the parameters. The walkers that are lost, meaning that they are outside the primary distribution of walkers, are ``trimmed" and not considered when determining parameter uncertainties.

\subsection{Model Fitting}\label{subsec:fitting}

In this paper, we fit models to the visibilities from ALMA, and combined visibility data from ALMA and the VLA to find the best fitting models of the continuum from HOPS 383. We do not include the visibility data from the SMA due to their lower signal-to-noise. The emcee code uses parallelization to spread the model calculations to a large number of cores, significantly speeding up the computation of the models. 

\subsection{Model Results}\label{subsec:models}
Our continuum dust models simultaneously reproduce the visibilities and the broadband SED for HOPS 383. We list the best fitting parameter values for our disk models of HOPS 383 in Table \ref{table:parameters}, and show the different models compared with the data in Figures \ref{fig:alma}, \ref{fig:alma-tapered}, \ref{fig:vla}, and \ref{fig:vla-tapered}. It is important to note that the models only directly constrain the dust masses in the disk and envelope. Thus the actual values constrained by the models are dust-only. In Table \ref{table:parameters}, however, we list the total mass using the canonical gas-to-dust ratio of 100. While we model the envelopes, their properties are expected to be poorly constrained due to the lack of short baseline data. Envelope mass and radii are expected to be highly uncertain. We discuss the results of our models below.

\subsubsection{HOPS 383 ALMA Model}
Although the envelope masses (0.38 $M_{\odot}$ and 1.1 $M_{\odot}$) and radii (5700 au and 12000 au) differ between the truncated and tapered model, the radii, and therefore the masses, are not well constrained and they are within the 1$\sigma$ uncertainties. Both models have a protostellar disk mass of $\approx$0.02 $M_{\odot}$ and independently find the same inclination ($i \approx 51^{\circ}$) and position angle ($P.A. \approx 49^{\circ}$). The model inclinations are also in good agreement with the estimate made from the deconvolved source size in Section \ref{subsec:continuum}. The tapered model has a smaller R$_{\text{disk}}$ than the truncated model, as expected. The truncated disk radius is similar to the disk radius we measured in Section \ref{subsec:continuum}. In both cases, the source is surrounded by a compact disk which is embedded within a large envelope which is more massive than the disk. 

The luminosities of both models are very similar with $L \sim 5$ $L_{\odot}$. The previous model of HOPS 383 by \citet{Safron15} using \textit{Spitzer}, \textit{Herschel}, and APEX photometry found luminosities of 6 $L_{\odot}< L_{bol} <$ 14 $L_{\odot}$, and inclination of 41$^{\circ} < i < 63^{\circ}$. Also, \citet{Furlan16} found $L_{bol}$ of 7.8 $L_{\odot}$. Our models have a consistent inclination but slightly lower luminosities. We believe that the SED fits we obtained from the models are not as optimal, underestimating the 100 $\mu$m and 160 $\mu$m region which results in a overall lower luminosity compared to luminosities obtained by \citet{Safron15} and \citet{Furlan16}.

\subsubsection{HOPS 383 ALMA+VLA Model}
The ALMA+VLA model results find massive disks of $M \sim 0.33 M_{\odot}$  and $M \sim 0.21 M_{\odot}$ for the truncated and the tapered cases, respectively. However, the disk radius of the truncated model is about 3 times larger than the tapered model, similar to the ALMA only case. The truncated disk model has a very similar envelope mass compared to the ALMA only models with $M_{env}$ = 0.32 $M_{\odot}$. However, the tapered disk model has very low envelope mass of 0.09 $M_{\odot}$. Both have smaller envelope radii ($\sim$2500 au) than the ALMA models, which is partially responsible for the low envelope masses. The truncated disk radius is similar to the disk radius we measured in Section \ref{subsec:continuum}. The luminosities are also fairly low with $\sim$4.0 $L_{\odot}$ for both models. The source appears to be less embedded than the ALMA only models with the disk more massive than the envelope. Similar to the ALMA models, our ALMA+VLA models also have lower luminosities compared to \citet{Safron15} and \citet{Furlan16}. The models do not fit short baselines as well as the ALMA only models and the fit to the far-IR and submillimiter SED is less optimal than the ALMA only models. 

\begin{figure*}[ht!]
\includegraphics[width=1.0\textwidth, trim={0 29.5cm 0 0},clip]{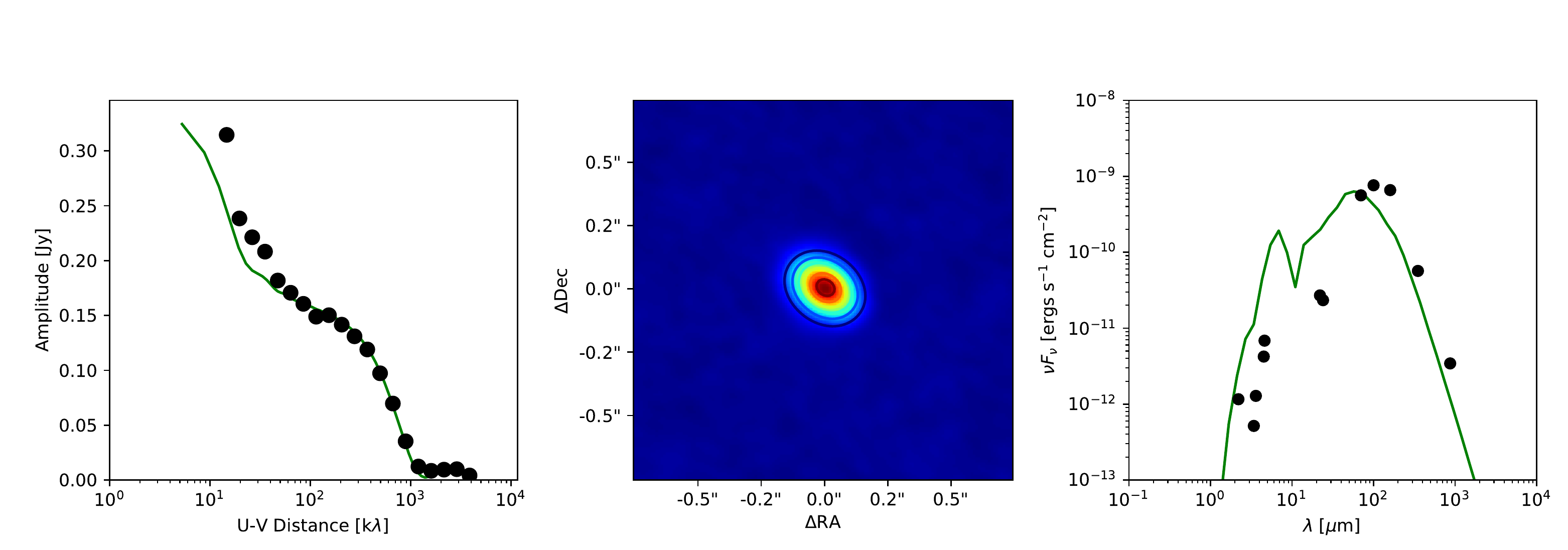}
\includegraphics[width=1.0\textwidth, trim={0 11cm 0 20cm},clip]{model_alma_vla.pdf}
\caption{Best fit model to HOPS 383 at 870 $\mu$m (top) and 8 mm (bottom) using the truncated disk. The 870 $\mu$m data comes from the ALMA and 8 mm comes from the VLA. The left panels show the circularly averaged visibility amplitudes of the data (black dots) compared to the model (green line). The middle panels show the data for ALMA (top) and the VLA (bottom) with the contours of the best fitting model overlaid. The right panel shows the model of the SED (green line) overlaid on the SED data from the literature. \label{fig:vla}}
\end{figure*}

\begin{figure*}[ht!]
\includegraphics[width=1.0\textwidth, trim={0 29.5cm 0 0},clip]{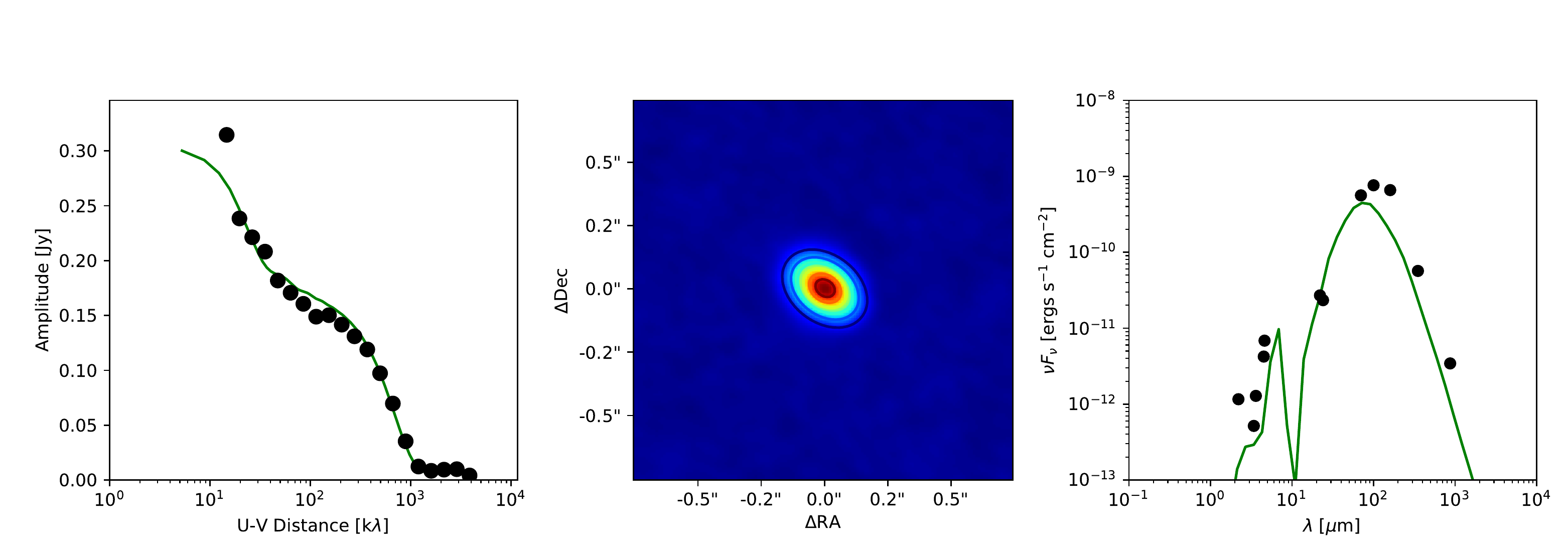}
\includegraphics[width=1.0\textwidth, trim={0 11cm 0 20cm},clip]{model_alma_vla_tapered.pdf}
\caption{Best fit model to HOPS 383 at 870 $\mu$m (top) and 8 mm (bottom) using the tapered disk. The 870 $\mu$m data comes from the ALMA and 8 mm comes from the VLA. The left panels show the circularly averaged visibility amplitudes of the data (black dots) compared to the model (green line). The middle panels show the data for ALMA (top) and the VLA (bottom) with the contours of the best fitting model overlaid. The right panel shows the model of the SED (green line) overlaid on the SED data from the literature. \label{fig:vla-tapered}}
\end{figure*}

Our models are not able to fully reproduce the visibilities from the VLA, without sacrificing the fit for the SED. This can be seen in the best-fit model of the VLA in Figures \ref{fig:vla} and \ref{fig:vla-tapered}, where the SED fit appears worse visually. The fitting to the visibility data from the VLA also does not account for a contribution from the free-free emission that was calculated in Subsection \ref{subsec:sed}. This mostly affects the fit to the baselines longer than 1000 k$\lambda$ since dust dominates on larger scales. This is probably why there is a small excess of emission in the longer baselines, and the points do not go to 0 like the ALMA visibilities (Figures \ref{fig:vla}, \ref{fig:vla-tapered}). This excess could also in part explain why the VLA models arrive at such higher disk masses. Model results will be further discussed in Section \ref{sec:discussion}.

\begin{figure*}[ht!]
\includegraphics[width=1.0\textwidth, trim={0 10cm 0 0},clip]{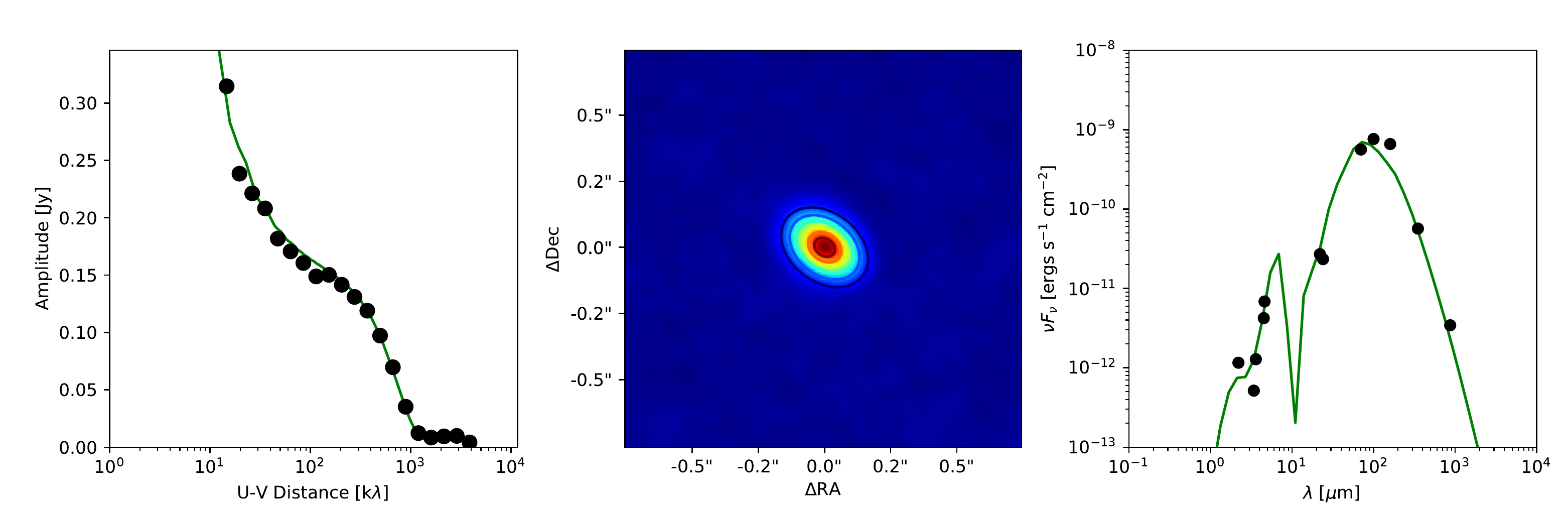}
\caption{Best fit model to HOPS 383 at 870 $\mu$m using the truncated disk. The left panel shows the circularly averaged visibility amplitudes of the data (black dots) compared to the model (green line) for ALMA (top) and the VLA (bottom). The middle panel shows the data with the contours of the best fitting model overlaid. The right panel shows the model of the SED (green line) overlaid on the SED data from the literature. \label{fig:alma}}
\end{figure*}

\begin{figure*}[ht!]
\includegraphics[width=1.0\textwidth, trim={0 10cm 0 0},clip]{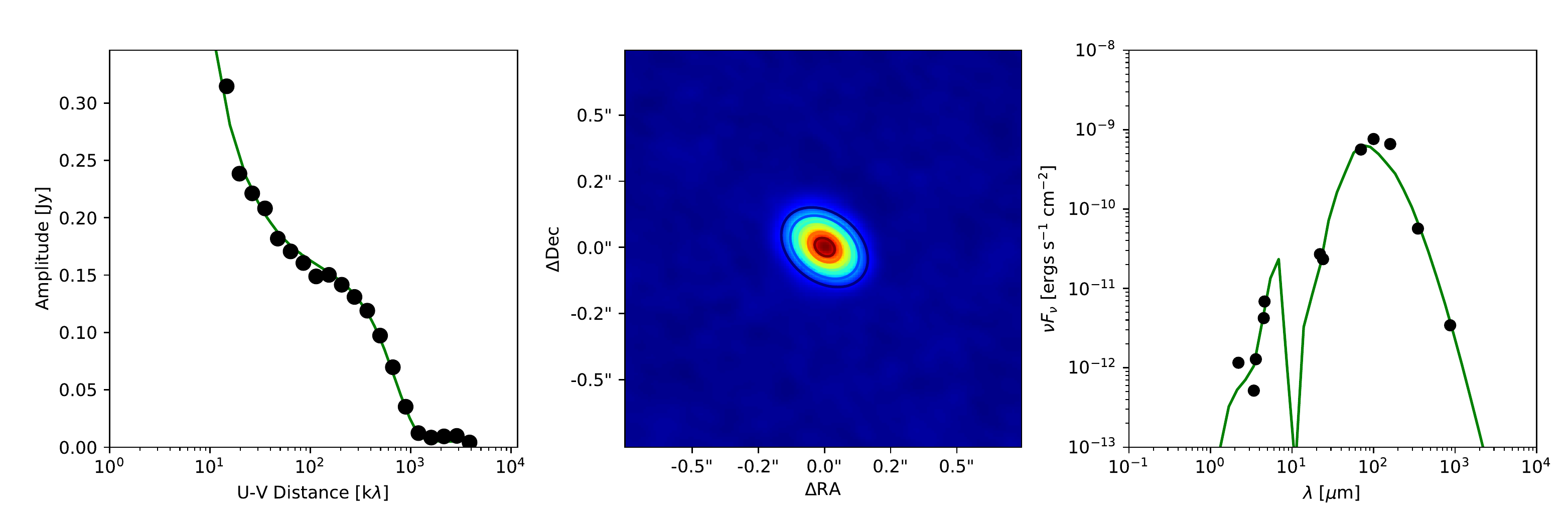}
\caption{Best fit model to HOPS 383 at 870 $\mu$m using the tapered disk. The left panel shows the circularly averaged visibility amplitudes of the data (black dots) compared to the model (green line) for ALMA (top) and the VLA (bottom). The middle panel shows the data with the contours of the best fitting model overlaid. The right panel shows the model of the SED (green line) overlaid on the SED data from the literature. \label{fig:alma-tapered}}
\end{figure*}

\begin{deluxetable*}{llccccBccccBccccBcccc}[t]
\centering
\tablecaption{Best fit model parameters for HOPS 383. \label{table:parameters}}
\tablenum{2}
\tablehead{
\colhead{Parameter (unit)} &
\colhead{Description} &
\multicolumn{2}{c}{ALMA} &
\multicolumn{2}{c}{ALMA+VLA} &\\
\colhead{} & \colhead{} & \colhead{Truncated} & \colhead{Tapered} & \colhead{Truncated} & \colhead{Tapered}
}
\startdata
$\beta$     &  Disk scale height exponent &        $0.90^{+0.07}_{-0.11}$ &           $0.61^{+0.12}_{-0.08}$ &        $0.77^{+0.03}_{-0.02}$ &        $0.80^{+0.04}_{-0.04}$ \\
$f_{cav}$    &  Outflow cavity scaling factor &       $0.13^{+0.05}_{-0.07}$ &           $0.09^{+0.03}_{-0.04}$ &        $0.15^{+0.03}_{-0.04}$ &        $0.16^{+0.06}_{-0.06}$ \\
$\gamma$      &  Surface density exponent   &     $0.65^{+0.11}_{-0.10}$ &          $-0.09^{+0.14}_{-0.09}$ &        $1.96^{+0.03}_{-0.05}$ &        $0.61^{+0.04}_{-0.05}$ \\
$h_{0} (\text{au})$   &  Scale height at 1 au &       $0.04^{+0.03}_{-0.02}$ &           $0.05^{+0.02}_{-0.02}$ &        $0.09^{+0.01}_{-0.01}$ &        $0.05^{+0.02}_{-0.01}$ \\
$i (^{\circ})$  & Inclination &          $51.1^{+ 0.3}_{- 0.3}$ &           $51.5^{+ 0.4}_{- 0.3}$ &        $50.9^{+ 1.0}_{- 0.7}$ &        $51.1^{+ 0.5}_{- 0.6}$ \\
$\xi$    & Cavity shape power law exponent    &    $1.21^{+0.05}_{-0.03}$ &           $1.22^{+0.05}_{-0.03}$ &        $0.83^{+0.13}_{-0.05}$ &        $1.10^{+0.03}_{-0.08}$ \\
$L_* (L_{\odot})$     &  Stellar luminosity &     $  5.0^{+  0.3}_{-  0.4}$ &        $  4.9^{+  0.3}_{-  0.3}$ &     $  4.7^{+  0.7}_{-  0.4}$ &     $  3.9^{+  0.2}_{-  0.1}$ \\
$M_{disk} (M_{\odot})$      &  Disk mass &     $0.026^{+0.004}_{-0.003}$ &        $0.020^{+0.005}_{-0.004}$ &     $0.330^{+0.025}_{-0.028}$ &     $0.214^{+0.011}_{-0.019}$ \\
$M_{env} (M_{\odot})$       &  Envelope mass &     $0.380^{+0.228}_{-0.138}$ &        $1.111^{+0.602}_{-0.340}$ &     $0.320^{+0.086}_{-0.055}$ &     $0.091^{+0.017}_{-0.010}$ \\
$R_{disk} (\text{au})$     &  Disk radius &     $64.96^{+ 1.28}_{- 1.20}$ &        $38.33^{+ 1.35}_{- 2.43}$ &     $57.67^{+ 0.99}_{- 1.05}$ &     $19.03^{+ 0.50}_{- 0.64}$ \\
$R_{env} (\text{au})$ & Envelope radius & $5857.16^{+2343.90}_{-1556.36}$ & $12543.30^{+4066.72}_{-2613.88}$ & $2687.71^{+530.65}_{-363.18}$ & $2423.51^{+278.53}_{-275.29}$ \\
$R_{in} (\text{au})$    &  Disk inner radius &     $ 1.61^{+ 0.28}_{- 0.24}$ &        $ 1.04^{+ 0.23}_{- 0.33}$ &     $ 0.95^{+ 0.14}_{- 0.11}$ &     $ 1.05^{+ 0.08}_{- 0.10}$ \\
$a_{max} ({\mu}\text{m})$ & Maximum grain size in disk &  $1116.57^{+650.31}_{-490.06}$ &    $1907.35^{+845.68}_{-544.45}$ &    $313.65^{+60.48}_{-87.77}$ &    $495.65^{+53.21}_{-71.69}$ \\
$p$     & Grain size distribution exponent &      $ 4.34^{+ 0.08}_{- 0.10}$ &        $ 3.81^{+ 0.12}_{- 0.06}$ &     $ 3.59^{+ 0.11}_{- 0.12}$ &     $ 3.81^{+ 0.09}_{- 0.22}$ \\
$P.A (^{\circ})$    &  Position angle &     $49.55^{+ 0.44}_{- 0.40}$ &        $49.43^{+ 0.32}_{- 0.42}$ &     $49.91^{+ 0.55}_{- 0.88}$ &     $49.35^{+ 0.77}_{- 0.70}$ \\
\enddata
\end{deluxetable*}

\section{Discussion}\label{sec:discussion}
\subsection{HCO$^+$ and N$_2$H$^+$}
The chemical effects of accretion bursts result from the change in temperature of the circumstellar gas and dust. In low temperature environment, $T \lesssim 20$ K, CO is frozen out on the dust grains and N$_2$H$^+$ is abundant in the cold regions where CO is frozen-out. However, as the dust temperature increases to T $\gtrsim$ 20 K due to the accretion burst, CO begins to sublimate off the dust grains. The CO evaporation rate is exponentially proportional to the temperature of the dust grains. Hence, the evaporation is effectively instantaneous. Consequently, N$_2$H$^+$ gets destroyed by proton exchange with the evaporating CO molecules forming HCO$^+$. This brings about a significant change on the radial abundance of these molecular species before and during bursts (\citealt{Visser12}; \citealt{Visser15}).

\begin{figure}[ht!]
\includegraphics[width=1.0\linewidth]{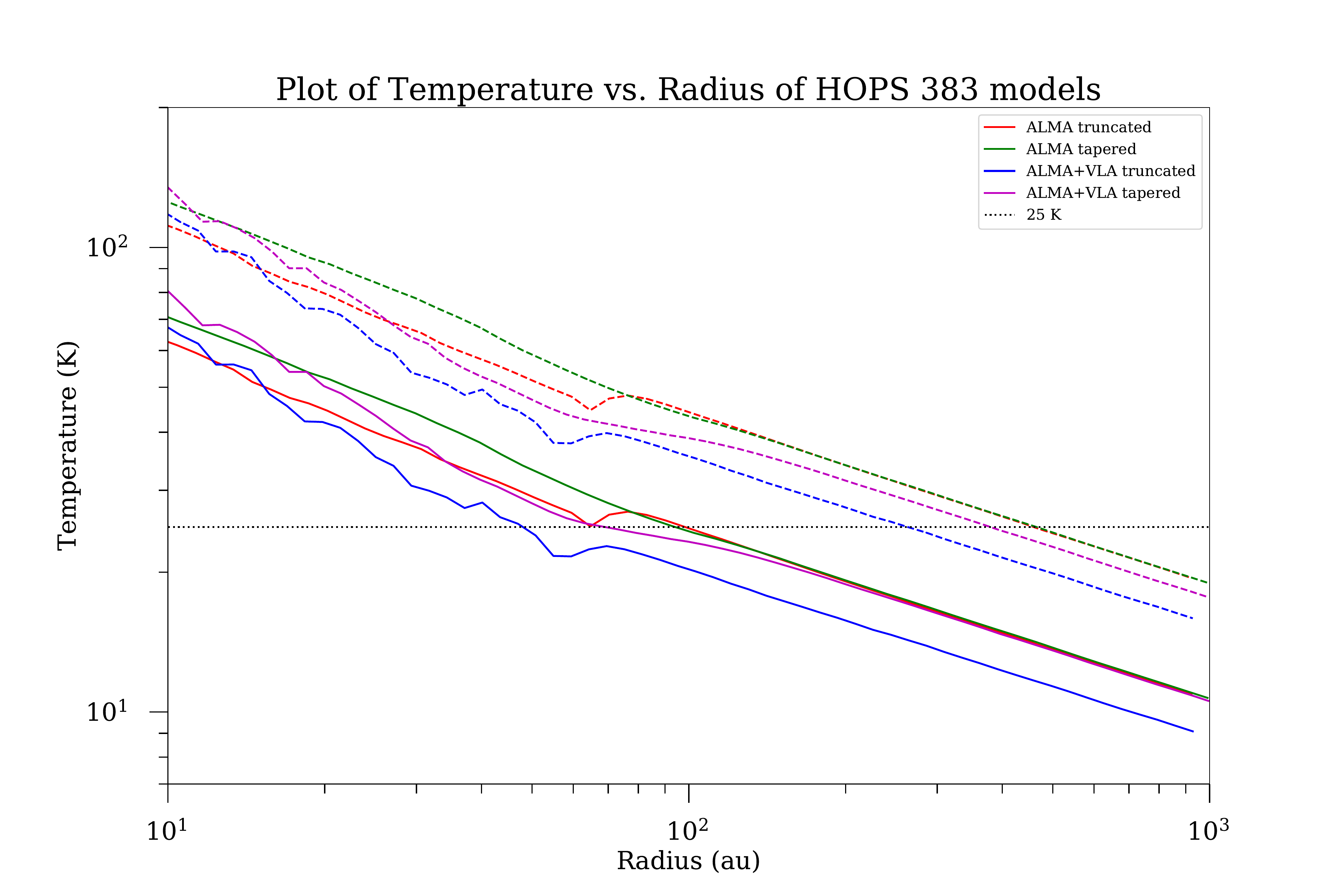}
\caption{Temperature as a function of radius for the different HOPS 383 models. The dashed lines represent the temperature values at the current luminosity derived from the models and the dashed lines represent the temperatures values prior to the outburst obtained by scaling the temperature by a factor of $\sqrt[4]{0.5 L_{\odot}/L_{*}}$, where 0.5 $L_{\odot}$ is taken to be the luminosity of HOPS 383 before the outburst. The dotted horizontal line marks the temperature value of 25 K. Currently the models have 25 K limit around 250 - 550 au. Before the outburst, the models have 25 K limit around 50 - 100 au. \label{fig:temp_models}}
\end{figure}

This effect can be seen in Figure \ref{fig:line}. At the central region near the protostar, where the temperature increase is the highest, we see that HCO$^+$ is peaked whereas N$_2$H$^+$ is reduced. Instead, N$_2$H$^+$ is double-peaked surrounding the protostar position. Such depression in N$_2$H$^+$ near the protostar is seen in non-bursting protostars of luminosity greater than 1 L$_{\odot}$ as well (e.g. \citealt{Hsieh19}; \citealt{Jorgensen04}; \citealt{Matthews06}; \citealt{Tobin11}). However, the pre-outburst luminosity of HOPS 383 is estimated to be around 0.2 - 0.5 L$_{\odot}$, which might not have heated the envelope to a level where the CO could sublimate out to such a large radius. We test this by looking at the radial temperature profiles of the models in Figure \ref{fig:temp_models}. We see that currently, the ALMA only and the ALMA+VLA models have 25 K temperature limit at $\sim$550 au radius and $\sim$250 - 400 au radius, respectively. These values are in agreement with the region where N$_2$H$^+$ is depressed in Figure \ref{fig:line}. Before the outburst, the models have the 25 K temperature limit at a radius around 50 - 100 au, which suggests that the depression in N$_2$H$^+$ would be closer to the protostar than it is currently situated.

Another possible alternative is that if the outflow knots discussed in Section \ref{subsubsec:knots} does in fact represent a previous burst about 215 yr ago, the observed structure of N$_2$H$^+$ could be a chemical feature that may represent the cumulative effect of multiple outbursts, rather than the latest one. It may not have been long enough since the last outburst for the changes in N$_2$H$^+$ to propagate to where we can resolve it in our observations. Nevertheless, in both cases, these observations are in good agreement with the models of \citet{Visser12}. However, it is important to note that while the models of \citet{Visser12} suggest that N$_2$H$^+$ should only be present outside R$\sim$500 AU, outside the region with abundant HCO$^+$, we do see spatial overlap between the HCO$^+$ and N$_2$H$^+$ emission. This is due to the longer destruction timescale of N$_2$H$^+$ compared to the instantaneous CO evaporation. The molecular destruction of N$_2$H$^+$ takes roughly 50 - 100 yr. Hence, there simply has not been enough time for the destruction of N$_2$H$^+$ in this region since the outburst in $\sim$2006.

Once the accretion burst is over and the protostar enters its quiescent phase, the temperature on the grains begins to drop. The evaporated CO eventually does freeze out onto the dust grains. However, the freeze out rate is significantly slower than the evaporation rate. The CO freeze out rate, given by the equation,
\begin{equation}
\tau_{\text{fr}}=1\times10^{4}\sqrt{\frac{\text{10 K}}{\text{T}_{\text{g}}}}\frac{10^{6}\text{ cm}^{-3}}{n(\text{H}_{2})},
\end{equation}
depends on the collision rate between molecules and the grains (\citealt{Charnley01}). This means that for a H$_2$ density of $10^6$ cm$^{-3}$ and gas temperature of 10 K, the freeze-out timescale of CO is $10^4$ yrs, longer than the $\sim$1000 yr interval between bursts inferred by \citet{Fischer19}. Thus, even after protostar enters its quiescent phase, the effects of the outburst are apparent for quite some time, explaining the extent of C$^{18}$O emission in the envelopes around protostars (\citealt{Jorgensen15}; \citealt{Frimann17}). This has the effect that N$_2$H$^+$ could continue to be destroyed out to larger radii long after the outburst.

\subsection{Long Term Chemical Effects of Accretion Burst}

In addition to N$_2$H$^+$ and HCO$^+$ mentioned above, the accretion burst also impacts the chemistry of other molecules. Various models have created to study the impact of outbursts in embedded protostars (e.g., \citealt{Jorgensen15}; \citealt{Rab17}; \citealt{Molyarova18}). These models can be used as guidelines on how different chemical species surrounding HOPS 383 are expected to evolve. \citet{Rab17} found that C$^{18}$O ($J = 2 \rightarrow 1$) stays in the gas phase for 10$^3$-10$^4$ yr and its post-burst emission can be fitted by a two Gaussian fit. Since our C$^{18}$O observations using SMA were contaminated, we refrain from conducting the Gaussian fits of our data. Likewise, C$^{18}$O freezes from the inside-out as the freeze-out timescale of $C^{18}$O is a function of the gas density, creating a gap in the intensity of C$^{18}$O (\citealt{Rab17}). This gap becomes visible by 1000 yr post-outburst and is visible as an X-shaped emission pattern for inclined targets such as HOPS 383.

Similarly, \citet{Molyarova18} studied post-burst chemical signature on the disk, in the absence of an embedded envelope. Particularly, they found that formaldehyde (H$_2$CO) grows by 4-6 orders of magnitude during the outburst and stays in the gas-phase for several decades as it freezes-out. We did observe H$_2$CO centered around the protostar with the SMA, however, we cannot determine whether the emission is primarily coming from the disk or inner envelope at the resolution of our observations. Nevertheless, the effect of an outburst on H$_2$CO in the inner envelope could be similar to the disk. \citet{Taquet16}, \citet{Molyarova18}, and \citet{Wiebe19} also suggest that observation of organic molecules such as CH$_3$OH and more complex species can be indicative of multiple outbursts. However, we did not detect any of the CH$_3$OH lines that were contained in our spectral windows. These lines are typically only detected for higher-luminosity protostars due to the excitation temperatures greater than 100 K (\citealt{Tobin19}).

\subsection{Flux Density Derived Toomre Q}
The observed flux density from ALMA gives us disk masses of 0.082 - 0.043 $M_{\odot}$ for adopted T$_{dust}$ = 30 K - 50 K, respectively.  Similarly the VLA masses range 0.1 to 0.06 $M_{\odot}$ for the same temperatures. We focus on the ALMA masses because the appropriate dust opacity at 9 mm for the VLA observations is uncertain. With these masses we can calculate Toomre's $Q$ parameter to assess the stability of the rotating disk of HOPS 383 against gravitational collapse. It calculates the ratio of rotation shear force and thermal pressure to the self-gravity of the disk,
\begin{equation} \label{eq:Q}
Q = \frac{c_{s}\Omega}{\pi G \Sigma},
\end{equation}
where $c_{s}$ is the sound speed, $\Omega$ is the differential rotation value of a Keplerian disk, $G$ is the gravitational constant, and $\Sigma$ is the surface density. The stability of the disk requires $Q > 1$, and $Q < 1$ indicates that the disk is gravitationally unstable and may be prone to fragmentation.

We can rewrite this equation (\citealt{Kratter16}; \citealt{Tobin16}) in the form,

\begin{equation}
Q \approx 2 \frac{M_{*} H}{M_{d} R},
\end{equation}
where, $M_*$ is the mass of the protostar, $H = c_{s}/{\Omega}$, $M_d$ is the mass of the disk and $R$ is the radius of the disk at which the $c_s$ is measured. Using a typical protostar mass of 0.5 M$_{\odot}$ (\citealt{Kenyon95}), we find that $Q = 1.67$ and 0.87 for M$_{\text{disk}} = 0.043 M_{\odot}$ and 0.082 $M_{\odot}$, respectively for the disk masses obtained from the ALMA flux density. Thus, at current estimates of $T_{dust} = 50$ K, $Q$ is $>$ 1 and the disk is being stabilized, but prior to the outburst when $T_{dust}$ was more likely $\sim$30 K, Q would be lower and closer to instability by a factor of $(50/30)^{0.5}$, as $Q$ scales by $\sqrt{T}$. Furthermore, with the likely underestimate of disk mass, Q would be lower as well. 

Likewise, since the pre-outburst luminosity was not well constrained, but estimated to be 0.2 L$_{\odot}$, the disk could have been even colder. This would lead toward a lower value of Q. Thus, while we cannot absolutely confirm that the disk is or was unstable, the disk appears to have sufficient mass to make gravitational instability a very plausible outburst mechanism for HOPS 383.

\subsection{Discussion of Best-fit Model Parameters}
Except for the ALMA tapered disk model, all other models have positive $\gamma$, i.e. decreasing surface density with increasing radius, as expected. The negative $\gamma$ for the ALMA tapered model can be attributed to the optically thick dust emission at 870 $\mu$m, which makes it difficult to constrain $\gamma$ very well. The emission at 8 mm is more optically thin and traces the dust in the disk more closely than at 870 $\mu$m.

A curious result of our models is the significantly higher disk mass of ALMA+VLA models compared to the ALMA only models. The SED fit to the ALMA+VLA models is noticeably worse than the ALMA only models, not fitting the far-IR and sub millimeter flux density that trace the bulk of the luminosity and envelope emission. This is likely because the ALMA+VLA models do not fit the envelope well. Similarly, the higher disk mass of the ALMA+VLA models could result from the 8 mm emission tracing the dust in the disk that is more optically thin compared to the 870 $\mu$m emission. In addition, the VLA data that we used for modeling contains a free-free contribution as well, which can further increase the estimate the disk mass.

The disk radius follows a similar pattern across the models, the tapered disk models have lower disk radii than the truncated models. It is difficult to properly define the disk radii for the tapered disk models as the disk does not end at the critical radii, $R_c$, like in the truncated models. Instead, at $R_c$ density of the disk starts to fall-off exponentially. Since the emission is likely optically thick, adding much more mass in the disk does not make it any brighter at short $\lambda$ (i.e., 870 $\mu$m). So we can have a very massive disk with a small $R_c$ where the disk density starts to fall-off, but it can still appear well-resolved out to the larger radii in the tapered models. For this reason, it is difficult to interpret radius of the tapered models compared to the truncated models. The radius of the disk where we see emission is better constrained by the truncated models. For this reason, we favor the truncated models values of disk radius. The ALMA only models have larger dust grains with $a_{max} >$ 1000  $\mu$m compared to the ALMA+VLA models. The dust in ALMA+VLA models have similar $a_{max} <  500 \mu$m with power-law grain size distributions of 3.59 for the truncated model and 3.81 for the tapered model. The larger dust grains of the ALMA only models might also contribute to the lower disk masses fitted for the ALMA only models.

The envelope mass and radius of the ALMA+VLA tapered model are significantly lower than its ALMA only counterparts. This has mostly to do with the number of points available for the fitting in the visibilities. The shorter uv-distances of the visibilities, specifically distances $< 100$ k$\lambda$ represent the envelope while the longer uv-distances represent the disk (see Figures \ref{fig:alma}, \ref{fig:alma-tapered}, \ref{fig:vla}, \ref{fig:vla-tapered}; left plots). There are only two points at less than 20 k$\lambda$ uv distance. Thus, our models will be able to best fit the inner envelope and disk. Likewise, the visibilities from the VLA have fewer points for fitting the envelope compared to the visibilities from the ALMA, reducing their weight in the fit. The envelope is expected to be composed of smaller dust and as such should have very few centimeter-sized dust grains. This means we get very low emission at the longer VLA wavelength from the envelope. The low envelope radii of the ALMA+VLA model also contributes to the lower envelope mass compared to the ALMA only model. However, we are most concerned in fitting the disk portion of the protostar, so that we can constrain the parameters of the disk and we are able to do that fairly well; even though we do not fit the envelope ideally.

\subsection{Toomre $Q$ vs. Radius}
\begin{figure*}[ht!]
\includegraphics[width=0.49\textwidth]{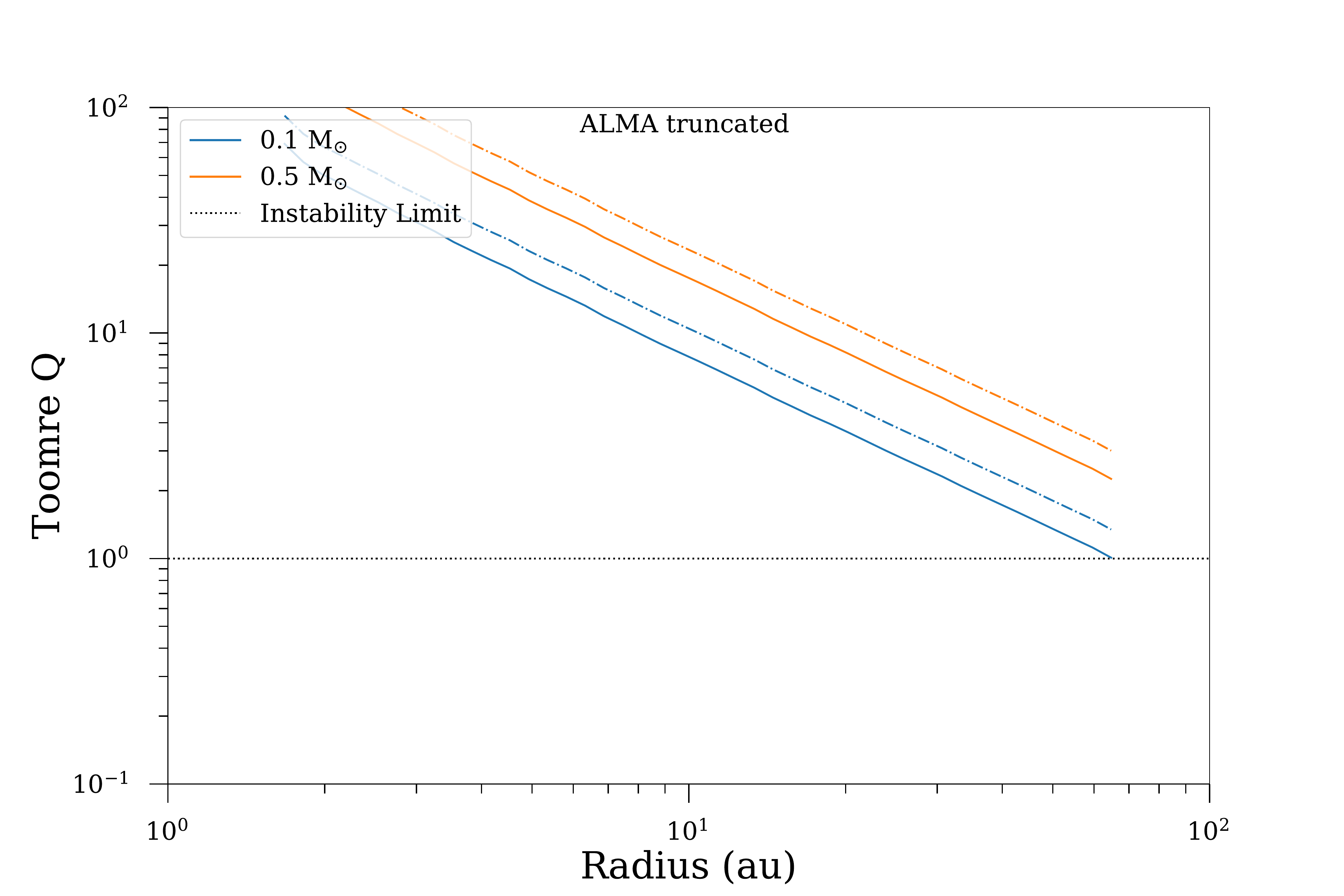}
\includegraphics[width=0.49\textwidth]{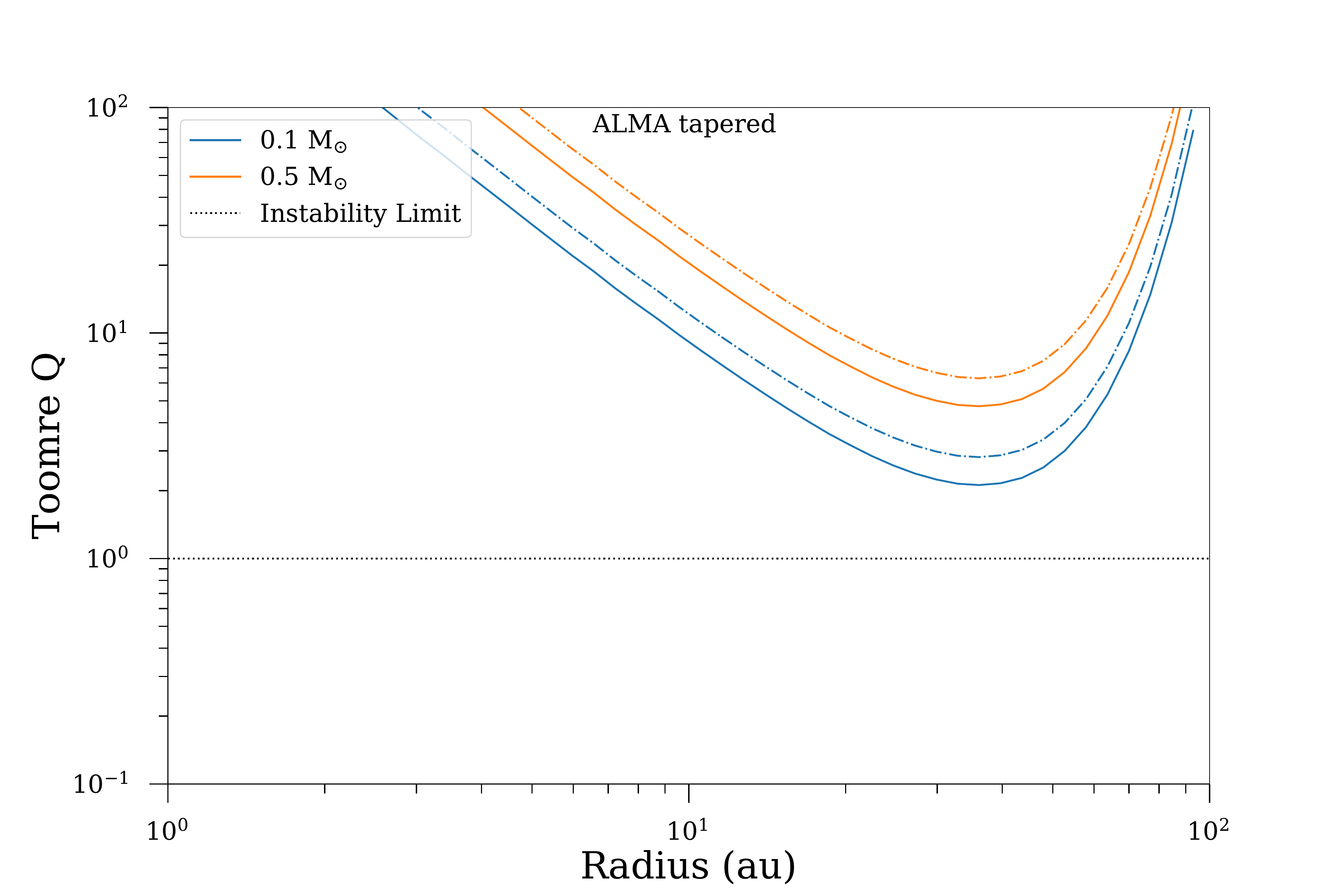}
\caption{Toomre's $Q$ as a function of radius for the disk of HOPS 383 using the ALMA only models for different protostar masses. The dashed lines represent the current $Q$ values and the solid lines repersent the Q values prior to the outburst obtained by scaling the temperature by a factor of $\sqrt[4]{0.5 L_{\odot}/L_{*}}$, where 0.5 $L_{\odot}$ is taken to be the luminosity of HOPS 383 before the outburst. The dotted horizontal line marks the value of $Q = 1$. Currently, the disk seems to be gravitationally stable for most cases. The truncated model seems to have been unstable around its critical radius. Although the tapered model has $Q > 1$ for all the cases besides 0.1 M$_{\odot}$ before the outburst too, the values are very close to 1, suggesting that the disk may have been gravitationally unstable before, causing the accretion burst. \label{fig:q-alma}}
\end{figure*}

The RADMC-3D synthetic models of HOPS 383 continuum in Section \ref{sec:model} have radial profiles of surface density and disk temperature embedded in them. Extracting these, we can use Equation \ref{eq:Q} to plot Toomre's $Q$ as a function of radius. Figure \ref{fig:q-alma} shows the $Q$ as a function of the radius using the data from ALMA only models whereas Figure \ref{fig:q-vla} does the same using the data from ALMA+VLA models. We assume the disk around HOPS 383 to be Keplerian since there is no evidence pointing otherwise. However, we do not know the exact mass of the central protostar, which changes the value of $\Omega$. Therefore we plot the $Q$ with different protostar masses of 0.1 M$_{\odot}$ to 0.5 M$_{\odot}$.

We assume a dust-to-gas mass ratio of 1:100 for our calculation. We take an average temperature of a few slices above the mid-plane for a more consistent temperature profile from the models. The dashed lines in the figures represent the current $Q$ values as a function of the disk radius. However, we need to determine $Q$ before the outburst to determine if GI was a viable mechanism for the outburst. To do this, we scale the disk temperature by $\sqrt[4]{L_{*, pre}/L_{*, burst}}$ based on the fact that following the outburst, the temperatures (and consequently $c_s$) are larger. We use the model derived luminosities for L$_{*, burst}$ and take L$_{*, pre}$ = 0.5 L$_{\odot}$; this is slightly larger than the estimated pre-burst luminosity from \citet{Safron15}. The solid lines in Figure \ref{fig:q-alma} and \ref{fig:q-vla} represent the Q values prior to the outburst.

The disk around HOPS 383 appears to be stable with $Q > 1$ for the ALMA only models. However, all these values are quite close to critical value of $Q = 1$, so the disk may have been susceptible to GI at $\sim$20 - 60 au, in the ALMA only models. The disks for the ALMA+VLA models appears to be gravitationally unstable around the critical radius in both cases. Furthermore, our millimeter observations are likely not sensitive to all the mass in the disk, which suggests that the disk is likely more massive than our ALMA only results. The higher the disk mass, the more likely it is to be gravitationally unstable. If the cause of the outburst is from a clump being accreted onto the protostar, the clump was most likely produced by GI. Hence, GI appears to be a viable mechanism for the accretion outburst in HOPS 383. 

\subsection{Estimating Free-Free Subtracted Disk Properties}
\begin{figure*}[ht!]
\includegraphics[width=0.5\textwidth]{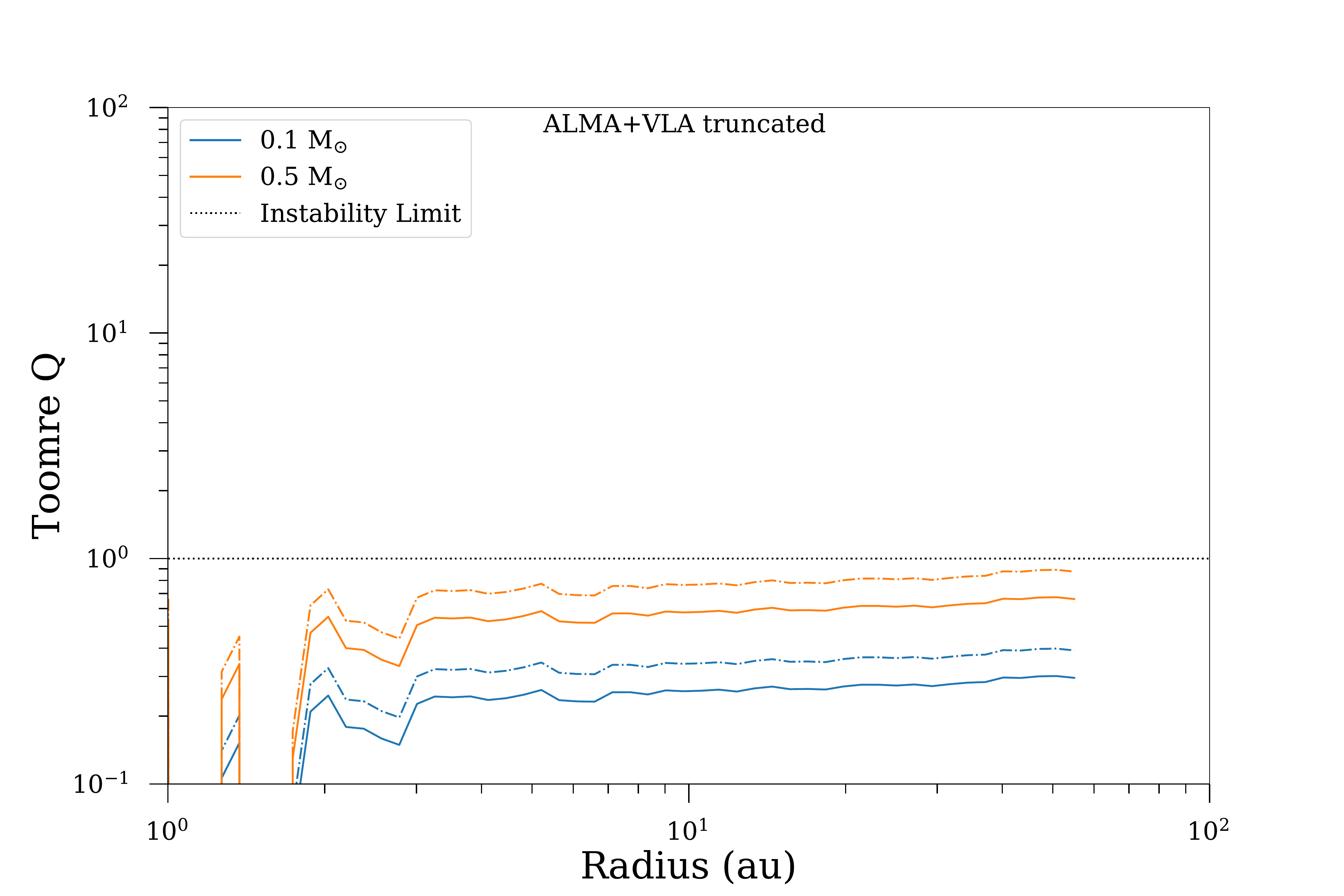}
\includegraphics[width=0.5\textwidth]{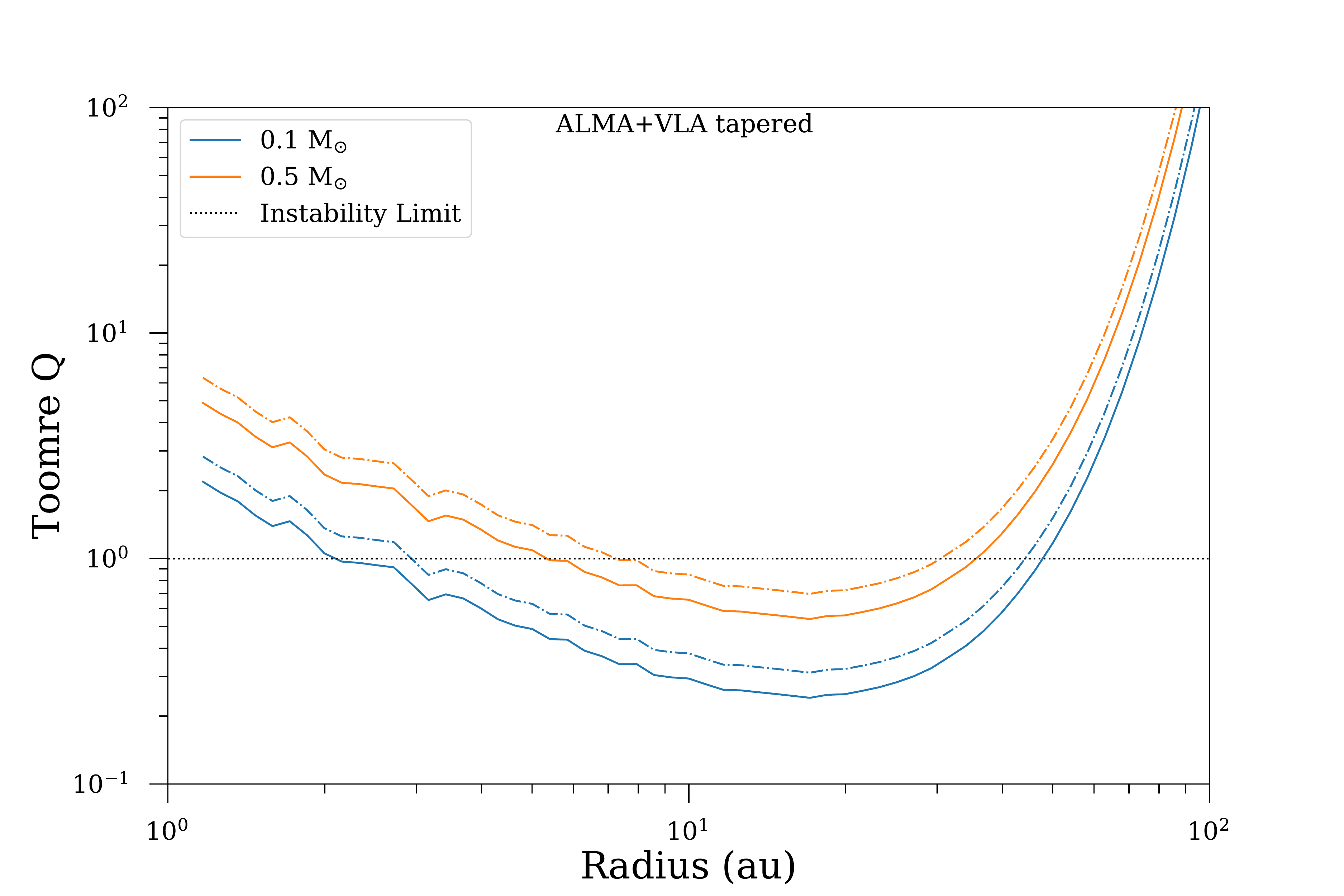}
\caption{Toomre's $Q$ as a function of radius for the disk of HOPS 383 using the ALMA+VLA models for different protostar masses. The different lines reperesent the same as Figure \ref{fig:q-alma}. Unlike the ALMA only models, the disk seems to currently be marginally unstable at the critical radius, $R_c$ in most cases. Prior to the outburst, we see that the disk is gravitationally unstable for all the cases, suggesting GI was at least partly responsible for the accretion burst. \label{fig:q-vla}}
\end{figure*}

For the ALMA+VLA models, the ratio of free-free emission to total flux density is 234 $\mu$Jy/411 $\mu$Jy $\approx 0.57$. Since the ALMA visibilities are not strongly affected by the free-free emission and can be fit with much lower disk masses, we can scale the ALMA+VLA disk masses by 0.43 (1-0.57) to estimate the mass without the free-free emission. Using the values from Table \ref{table:parameters}, this gives us disk masses of $\sim$0.14 M$_{\odot}$ and $\sim$0.092 M$_{\odot}$ for the truncated and the tapered ALMA+VLA models, respectively. The lower disk mass would decrease the surface density of the disk and consequently change the Toomre's $Q$ value as a function of the disk radius as $M_d \propto \Sigma R_d^2$. Thus, the $Q$ values for the free-free subtracted cases would be about 2.3 times higher than shown in Figure \ref{fig:q-vla}. However, to obtain realistic parameters, we need to run models with the free-free contribution incorporated from the VLA visibilities. We are currently running these models and will present the findings in a later paper.

\section{Conclusions}\label{sec:conclusion}
We have characterized the first known outbursting Class 0 protostar. We have examined the chemical signatures expected for an outburst and the disk to characterize the origin of the outburst with spatial resolutions of 1600 to 500 au. The effects of the outburst are seen on the chemistry of the highly embedded protostellar system of HOPS 383. The two peaks in the  N$_2$H$^+$ intensity straddle the position of the protostar; although of marginal significance, this morphology suggests that this molecule is being destroyed by reactions with the evaporating CO molecules, forming HCO$^+$. The HCO$^+$ emission is concentrated around the protostar position, where N$_2$H$^+$ is reduced (\citealt{Jorgensen04}). Our models suggest that the depression in N$_2$H$^+$ indicates the region where CO evaporated after the outburst as suggested by \citet{Visser12}. Since the freeze-out of CO takes about 10$^4$ yr, N$_2$H$^+$ in the inner regions of the envelope should continue to react with CO to larger radii, significantly changing the radial abundances of these molecules. We also observe emission from N$_2$D$^+$ and H$^{13}$CO$^+$ which do not peak as strongly but follow the spatial distributions of N$_2$H$^+$ and HCO$^+$, as expected from an outburst.

The ALMA and the VLA continuum is observed with $\sim$40 au resolution. The ALMA observations resolve the dust continuum at 870 $\mu$m, revealing a protostellar disk orthogonal to the outflow. The VLA observation at 9 mm is resolved but is much more compact than the ALMA observations. We estimate the disk mass from the ALMA and the VLA continuum to be $\sim$0.043 M$_{\odot}$ and $\sim$0.15 M$_{\odot}$, respectively for a typical dust temperature of 50 K after the outburst, as indicated by its luminosity of 7.8 L$_{\odot}$. Using the ALMA disk mass and a typical protostar mass of 0.5 M$_{\odot}$, we calculate Toomre's $Q$, finding a disk becoming gravitationally stabilized with $Q$ = 1.67. Prior to the outburst, however, the lower luminosity would lead to a lower a lower disk temperature, making it likely to be gravitationally unstable. 

We fit  a radiative model of a disk and envelope heated by an internal protostar to the ALMA only visibilities and the ALMA+VLA visibilities. In both cases, the observed SED of the protostar was an additional constraint. We found that the ALMA visibilities can be well modeled with an envelope and disk. The best fit truncated disk model has a disk mass of 0.1 M$_{\odot}$ and a radius of $\sim$67 au, whereas the tapered model has a disk mass of 0.065 M$_{\odot}$ and a radius of $\sim$40 au. The ALMA+VLA models indicate $\sim$10$\times$ more disk mass than ALMA only models. We also examine Toomre's $Q$ as a function of the disk radius, finding that the disks of ALMA only models are currently gravitationally stable at most radii but could have been marginally unstable prior to the outburst around the critical radius. However, the ALMA+VLA models appear to be unstable, providing an indication that GI could be responsible for the outburst.

\acknowledgments

The authors thank the anonymous referee for a constructive report that improved the quality of the manuscript. RS and JTT acknowledge support from NSF Astronomy and Astrophysics grant \# AST-1814762. JKJ acknowledges support by the European Research Council (ERC) under the European Union's Horizon 2020 research programme through ERC Consolidator Grant ``S4F'' (grant agreement No~646908). The Submillimeter Array is a joint project between the Smithsonian Astrophysical Observatory and the Academia Sinica Institute of Astronomy and Astrophysics and is funded by the Smithsonian Institution and the Academia Sinica. The authors wish to recognize and acknowledge the very significant cultural role and reverence that the summit of Mauna Kea has always had within the indigenous Hawaiian community.  We are most fortunate to have the opportunity to conduct observations from this mountain. This paper makes use of the following ALMA data: ADS/JAO.ALMA\#2015.1.00041.S.ALMA is a partnership of ESO (representing its member states), NSF (USA) and NINS (Japan), together with NRC (Canada), MOST and ASIAA (Taiwan), and KASI (Republic of Korea), in cooperation with the Republic of Chile. The Joint ALMA Observatory is operated by ESO, AUI/NRAO and NAOJ. The National Radio Astronomy Observatory is a facility of the National Science Foundation operated under cooperative agreement by Associated Universities, Inc. [Some of] The computing for this project was performed at the OU Supercomputing Center for Education \& Research (OSCER) at the University of Oklahoma (OU).
\facilities{SMA, VLA, ALMA}
\software{pdspy (\citealt{Sheehan18}), CASA (\citealt{McMullin07}), RADMC-3D (\citealt{Dullemond12}), emcee (\citealt{Foreman13}), matplotlib (\citealt{Hunter07}), MIRIAD (\citealt{Sault95})}.

\end{document}